\documentstyle[12pt,graphicx]{article}                
\begin{document}
\baselineskip 18pt
\def\today{\ifcase\month\or
 January\or February\or March\or April\or May\or June\or
 July\or August\or September\or October\or November\or Decembe\fi
 \space\number\day, \number\year}
\def\thebibliography#1{\section*{References\markboth
 {References}{References}}\list
 {[\arabic{enumi}]}{\settowidth\labelwidth{[#1]}
 \leftmargin\labelwidth
 \advance\leftmargin\labelsep
 \usecounter{enumi}}
 \def\newblock{\hskip .11em plus .33em minus .07em}
 \sloppy
 \sfcode`\.=1000\relax}
\let\endthebibliography=\endlist
\def\lsim{\ ^<\llap{$_\sim$}\ }
\def\gsim{\ ^>\llap{$_\sim$}\ }
\def\r2{\sqrt 2}
\def\beq{\begin{equation}}
\def\eeq{\end{equation}}
\def\beqn{\begin{eqnarray}}
\def\eeqn{\end{eqnarray}}

\begin{titlepage}

\begin{center}
{\large {\bf Neutralino decay of MSSM neutral Higgs bosons}}\\
\vskip 0.5 true cm
\vspace{2cm}
 Tarek Ibrahim  
\vskip 0.5 true cm
\end{center}

\noindent
\begin{center}
{Department of  Physics, Faculty of Science,
Alexandria University, Egypt}\\
{and}\\
{Department of Physics, Northeastern University, Boston, MA 02115-5000, USA \footnote{Current address}}\\
\end{center} 
\vskip 1.0 true cm
\centerline{\bf Abstract}
\medskip
We compute the one loop corrected effective Lagrangian for the neutralino-neutralino-neutral Higgs interactions $\chi^{0}_{\ell} \chi^{0}_kH^{0}_m$. The analysis completes the previous analyses where similar corrections were computed for the $\bar{f} f H^{0}_m$ couplings, where $f$ stands for Standard Model quarks and leptons  and for
the chargino-chargino-neutral Higgs couplings
 $\chi^{+}_l \chi^{-}_kH^{0}_m$  within the minimal supersymmetric standard model MSSM. 
The effective one loop Lagrangian is then applied to the computation of the neutral Higgs decays. 
The sizes of the supersymmetric loop corrections of the neutral Higgs decay widths
into $\chi^{0}_{\ell} \chi^{0}_k$ (${\ell}=1,2,3,4$; $k=1,2,3,4$) are investigated and the supersymmetric
loop correction is found to be in the range of $\sim10\%$ in 
significant   regions  of the parameter space. By including the loop corrections
of the other decay channels $\bar{b} b$, $\bar{t} t$, $\bar{\tau} \tau$, $\bar{c} c$, 
and $\chi^{-}_i \chi^{+}_j$ ($i=1,2$; $j=1,2$), the corrections   to  branching ratios for
$H^{0}_m\rightarrow  \chi^{0}_{\ell} \chi^{0}_k$ can reach as high as $50\%$. 
 The effects of CP phases on the branching ratio are also investigated.
A 
discussion of the implications of the analysis for colliders is given.
\end{titlepage}
\section{INTRODUCTION}

The Higgs couplings to matter and gauge fields are of current interest
as they affect different phenomena which could be tested in low energy
processes \cite{carena1}. Recently calculations of the supersymmetric one loop
corrections to the Higgs boson couplings were given and their implications
for the neutral Higgs 
boson decays
into $\bar{b} b$, $\bar{t} t$, $\bar{\tau} \tau$, 
$\bar{c} c$ and $\chi^{-}_i \chi^{+}_j$ were analyzed \cite{me1}.
These decays are of great importance as they differ from the Higgs decay
predictions in the Higgs sector of the standard model. In this work we extend
the analysis to include the loop corrections of the $\chi^{0}_{\ell} \chi^{0}_kH^{0}_m$
couplings and the neutral Higgs decay into pairs of neutralinos. The complete 
analysis of the one loop corrected partial widths of the above channels allows one
to investigate also the effects of these corrections on the branching ratios
of different modes.

In this paper we include the effect of CP phases arising from the soft supersymmetric
breaking parameters. 
It is well known that large CP phases 
would induce electric dipole moments of the fermions in the theory. However these large
CP phases 
can be made compatible \cite{edm1,edma,edmb} with the severe experimental constraints
that exist on the electric dipole moments of the electron \cite{edm2}, of the neutron \cite{edm3}, and of the $Hg^{199}$ \cite{edm4}.
It is well known that if the phases are large they affect a variety of low energy phenomena \cite{rpm}. Some works in this direction have included the effects of CP phases on the neutral Higgs boson system. These phases induce mixings between the neutral CP even and the
CP odd Higgs  and can affect the decay of the neutral and charged Higgs into 
different modes \cite{Higgs}.

The current analysis of $\Delta {\cal{L}}_{\chi^{0}\chi^{0}H^{0}}$ and 
neutral Higgs decay into neutralinos is based on the effective Lagrangian
method where the couplings of the electroweak eigen states $H^{1}_1$ and 
 $H^{2}_2$ with neutralinos are radiatively corrected using the zero external
momentum approximation. The same technique has been used in calculating the 
effective Lagrangian and decays of $H^{0}_m$ into quarks and leptons \cite{carena1,babu,we1} and into chargino pairs \cite{me1}.
It has been used also in the analysis of the effective Lagrangian of charged Higgs with quarks \cite{carena1,two} and their decays into ${\bar{t}}b$ and $\nu_{\tau} \tau$
\cite{we2}
and into chargino $+$ neutralino \cite{we3}.
The neutral Higgs decays into neutralinos have been investigated before in the
CP conserving case \cite{eberl,ren}. 
 However, the analysis for the neutral Higgs 
decays into neutralinos, with one loop corrections, in the CP violating case where the neutral Higgs sector
is modified in couplings, spectrum and mixings, does not exist.
We evaluate the radiative corrections to the Higgs boson masses
and mixngs by using the effective potential approximation.
We include the corrections from the top and bottom quarks and squarks 
\cite{demir1}, from the chargino, the W and the charged 
Higgs sector \cite{we4} and from the neutralino, Z boson, and 
the neutral Higgs bosons \cite{we5}.
It is important to notice that the corrections to the Higgs effective
potential from the different sectors mentioned above are all one-loop corrections.
The corrections of the interaction $\Delta {\cal{L}}_{\chi^{0}\chi^{0}H^{0}}$
to be considered in this work are all one-loop level ones. So the analysis
presented here is a consistent one loop study. 

The outline of the rest of the paper is as follows: In Sec. 2 we compute the effective Lagrangian for the $\chi^{0}_{\ell} \chi^{0}_kH^{0}_m$ interaction. In Sec. 3 we give an analysis of the decay widths of the neutral Higgs bosons into neutralinos using the effective Lagrangian. In Sec. 4 we give a numerical analysis of the size of the loop effects on the partial decay widths and on the  branching ratios. In Sec. 5 we discuss the implications of the corrections considered here,
in the environment of the Large Hadron Collider LHC. Conclusions are given in Sec. 6.

\section{LOOP CORRECTIONS TO NEUTRAL HIGGS COUPLINGS}

The tree-level Lagrangian for $\chi^{0}_{\ell} \chi^{0}_kH^{0}$ interaction is
\beq
{\cal{L}}=\theta_{k{\ell}}\overline{\chi_{k}^{0}} P_L \chi^{0}_{\ell} H^{1*}_1+\tau_{k{\ell}}\overline{\chi_k^{0}} P_R \chi^{0}_{\ell} H^{2}_2+H.c.,
\label{i1}
\eeq
where $H^1_1$ and $H^2_2$ are the neutral states of the two Higgs isodoublets in the minimal supersymmetric standard model (MSSM), i.e.,
\beqn
(H_1)= \left(\matrix{H_1^1\cr
 H_1^2}\right),~~
(H_2)= \left(\matrix{H_2^1\cr
             H_2^2}\right)
\eeqn
and $\theta_{k{\ell}}=-gQ^{*'}_{k{\ell}}$ and 
$\tau_{k{\ell}}=gS^{'}_{{\ell}k}$ where
\beq
Q^{'}_{ij}=\frac{1}{\sqrt 2}[X^{*}_{3i}(X^{*}_{2j}-\tan \theta_W
X^{*}_{1j})]
\eeq
\beq
S^{'}_{ij}=\frac{1}{\sqrt 2}[X^{*}_{4j}(X^{*}_{2i}-\tan \theta_W
X^{*}_{1i})]
\eeq
The matrix elements $X$ are defined as
\beq
X^T M_{\chi^0} X=diag(m_{\chi^0_1}, m_{\chi^0_2}, m_{\chi^0_3}, m_{\chi^0_4})
\eeq
where $M_{\chi^0}$ is the $4\times 4$ neutralino mass matrix.

The loop corrections produce shifts in the couplings of Eq.~({\ref{i1}}) and the effective Lagrangian with loop corrected couplings is given by
\beqn
{\cal{L}}_{eff}=(\theta_{k{\ell}}+\delta \theta_{k{\ell}})\overline{\chi_k^{0}} P_L \chi^{0}_{\ell} H^{1*}_1+ \Delta \theta_{k{\ell}} \overline{\chi_k^{0}} P_L \chi^{0}_{\ell} H^{2}_2+\nonumber\\
~~(\tau_{k{\ell}}+\delta \tau_{k{\ell}})\overline{\chi_k^{0}} P_R \chi^{0}_{\ell} H^{2}_2+ \Delta \tau_{k{\ell}} \overline{\chi_k^{0}} P_R \chi^{0}_{\ell} H^{1*}_1
+H.c.
\label{t1}
\eeqn
In this work we calculate the loop corrections
$\delta \theta_{k{\ell}}$, $\Delta \tau_{k{\ell}}$, $\Delta \theta_{k{\ell}}$
and $\delta \tau_{k{\ell}}$ 
  using the zero external momentum approximation.

\subsection{Loop analysis of $\delta \theta_{k{\ell}}$ and  $\Delta \tau_{k{\ell}}$}
Contributions to $\delta \theta_{k{\ell}}$ and  $\Delta \tau_{k{\ell}}$ arise from the fourteen loop diagram of Fig. 1. We discuss now in detail the contribution of each of these diagrams. 
The basic integral that enters in the loop analysis is
\beq
J=\int \frac{d^4\ell}{(2\pi)^4}\frac{1}{(\ell^2-m^2_1+i\epsilon)
(\ell^2-m^2_2+i\epsilon)(\ell^2-m^2_3+i\epsilon)}
\eeq
where $m_1$, $m_2$ and $m_3$ are the masses of the particles running inside
the loops. This integral gives
\beq
J=\frac{i}{(4\pi)^2} f(m^2_1,m^2_2,m^2_3)
\eeq
where
\beqn
f(m^2_1,m^2_2,m^2_3)=\frac{1}{(m^2_1-m^2_3)} \frac{1}{(m^2_3-m^2_2)} \frac{1}{(m^2_1-m^2_2)}\nonumber\\
\times [m^2_2m^2_3\ln(\frac{m^2_2}{m^2_3})+
m^2_3m^2_1\ln(\frac{m^2_3}{m^2_1})+m^2_1m^2_2\ln(\frac{m^2_1}{m^2_2})]
\eeqn
and for the case of $m_2=m_3$, one finds
\beq
J=\frac{i}{(4\pi)^2} \frac{1}{(m^2_3-m^2_1)^2} [m^2_1\ln(\frac{m^2_3}{m^2_1})+m^2_1-m^2_3]
\eeq
We begin with the loop diagram of Fig. 1(i), part (a), which contributes the following to $\delta \theta_{k{\ell}}$ and  $\Delta \tau_{k{\ell}}$:
\beqn
\delta\theta^{(1)}_{k{\ell}}=-\sum_{i=1}^2\sum_{j=1}^2 
\frac{m_t}{8\pi^2}
F^{*}_{ji} (\alpha_{t{\ell}}D_{t1j}-\gamma_{t{\ell}}D_{t2j})\nonumber\\
\times (\beta^{*}_{tk}D^{*}_{t1i}+\alpha_{tk}D^*_{t2i}) f(m^2_t, m^2_{\tilde{t}_i}, m^2_{\tilde{t}_j})\nonumber\\
\Delta\tau^{(1)}_{k{\ell}}=-\sum_{i=1}^2\sum_{j=1}^2 
\frac{m_t}{8\pi^2}
F^{*}_{ji} (\beta_{t{\ell}}D_{t1j}+\alpha^{*}_{t{\ell}}D_{t2j})\nonumber\\
\times (\alpha^{*}_{tk}D^{*}_{t1i}-\gamma^{*}_{tk}D^*_{t2i}) f(m^2_t, m^2_{\tilde{t}_i}, m^2_{\tilde{t}_j})
\eeqn
where $F_{ji}$ is given by
\beqn
F_{ji}=-\frac{gM_Z}{\sqrt 2 \cos\theta_W}
((\frac{1}{2}-\frac{2}{3}\sin^2\theta_W)D_{t1j}^{*}D_{t1i}
+\frac{2}{3}\sin^2\theta_W D_{t2j}^{*}D_{t2i})\cos\beta \nonumber\\
+\frac{g m_t \mu}{\sqrt 2 m_W \sin\beta}
D_{t1j}^{*}D_{t2i}
\eeqn
The couplings $\alpha_{tk}$, $\beta_{tk}$ and $\gamma_{tk}$ are given by
\beqn
\alpha_{tk} =\frac{g m_tX_{4k}}{2m_W\sin\beta}\nonumber\\
\beta_{tk}=eQ_tX_{1k}^{'*} +\frac{g}{\cos\theta_W} X_{2k}^{'*}
(T_{3t}-Q_t\sin^2\theta_W)\nonumber\\
\gamma_{tk}=eQ_t X_{1k}'-\frac{gQ_t\sin^2\theta_W}{\cos\theta_W}
X_{2k}'
\eeqn
where $X'$'s are given by 
\beqn
X'_{1k}=X_{1k}\cos\theta_W +X_{2k}\sin\theta_W\nonumber\\
X'_{2k}=-X_{1k}\sin\theta_W +X_{2k}\cos\theta_W
\eeqn
The matrix elements $D_q$  are diagonalizing the squark mass$^2$ matrix as follows
\beqn
D^{+}_q M^2_{\tilde{q}} D_q =diag (m^2_{\tilde{q1}},
m^2_{\tilde{q2}})
\eeqn
Next for the loop Fig. 1(i), part(b), we find
\beqn
\delta\theta^{(2)}_{k{\ell}}=-\sum_{i=1}^2\sum_{j=1}^2 
\frac{m_b}{8\pi^2}
H^{*}_{ij} (\alpha_{b{\ell}}D_{b1j}-\gamma_{b{\ell}}D_{b2j})\nonumber\\
\times (\beta^{*}_{bk}D^{*}_{b1i}+\alpha_{bk}D^*_{b2i}) f(m^2_b, m^2_{\tilde{b}_i}, m^2_{\tilde{b}_j})\nonumber\\
\Delta\tau^{(2)}_{k{\ell}}=-\sum_{i=1}^2\sum_{j=1}^2 
\frac{m_b}{8\pi^2}
H^{*}_{ij} (\beta_{b{\ell}}D_{b1j}+\alpha^{*}_{b{\ell}}D_{b2j})\nonumber\\
\times (\alpha^{*}_{bk}D^{*}_{b1i}-\gamma^{*}_{bk}D^*_{b2i}) f(m^2_b, m^2_{\tilde{b}_i}, m^2_{\tilde{b}_j})
\eeqn
and $H_{ij}$ is given by
\beqn
H_{ij}=-\frac{gM_Z}{\sqrt 2 \cos\theta_W}
((-\frac{1}{2}+\frac{1}{3}\sin^2\theta_W)D_{b1i}^{*}D_{b1j}
-\frac{1}{3}\sin^2\theta_W D_{b2i}^{*}D_{b2j})\cos\beta \nonumber\\
-\frac{gm^2_b}{\sqrt 2 m_W \cos\beta}(D_{b1i}^{*}D_{b1j}+
D_{b2i}^{*}D_{b2j})-\frac{g m_b A_b}{\sqrt 2 m_W \cos\beta}
D_{b2i}^{*}D_{b1j}
\eeqn

For the loop of Fig. 1(ii), part(a), we find
\beqn
\delta\theta^{(3)}_{k{\ell}}=0\nonumber\\
\Delta\tau^{(3)}_{k{\ell}}=0
\eeqn

For the loop of Fig. 1(ii), part(b), we find
\beqn
\delta\theta^{(4)}_{k{\ell}}=0\nonumber\\
\Delta\tau^{(4)}_{k{\ell}}=\sum_{j=1}^2\frac{h_bm^2_b}{8\pi^2}
(\beta_{b{\ell}}D_{b1j}+\alpha^{*}_{b{\ell}}D_{b2j})\nonumber\\
\times (\alpha^{*}_{bk}D^{*}_{b1j}-\gamma^{*}_{bk}D^*_{b2j}) f(m^2_{\tilde{b}_j}, m^2_{b}, m^2_{b})
\eeqn
where $h_b$ is given by
\beq
h_{b}=\frac{gm_{b}}{\sqrt 2 m_W \cos\beta}
\eeq
 
For loop of Fig. 1(ii), part(c), we find
\beqn
\delta\theta^{(5)}_{k{\ell}}=-\sum_{i=1}^4\sum_{j=1}^4\frac{g^3}{2\pi^2\cos^2\theta_W}
Q^{'*}_{ij}R^{'''}_{kj}L^{'''}_{i{\ell}}\nonumber\\
\times m_{\chi^0_i}m_{\chi^0_j}f(m^2_{\chi^0_i}, m^2_{\chi^0_j}, m^2_{Z})\nonumber\\
\Delta\tau^{(5)}_{k{\ell}}=0
\eeqn
where the couplings  $L^{'''}_{ij}$ and $R^{'''}_{ij}$ are given
by
\beq
L^{'''}_{ij}=-R^{'''*}_{ij}=-\frac{1}{2}X^{*}_{3i}X_{3j}+\frac{1}{2}X^{*}_{4i}X_{4j}
\eeq

For loop of Fig. 1(ii), part(d), we find
\beqn
\delta\theta^{(6)}_{k{\ell}}=\sum_{i=1}^4\sum_{j=1}^4\sum_{n=1}^3\frac{g^3}{4\pi^2}
Q^{'*}_{ij}\nonumber\\
\{Q^{'*}_{i{\ell}}(Y_{n1}-iY_{n3}\sin\beta)-S^{'*}_{i{\ell}}(Y_{n2}-iY_{n3}\cos\beta)\}\nonumber\\
\times\{Q^{'*}_{kj}(Y_{n1}-iY_{n3}\sin\beta)-S^{'*}_{kj}(Y_{n2}-iY_{n3}\cos\beta)\}\nonumber\\
m_{\chi^0_i}m_{\chi^0_j}f(m^2_{\chi^0_i}, m^2_{\chi^0_j}, m^2_{H^0_n})\nonumber\\
\Delta\tau^{(6)}_{k{\ell}}=0
\eeqn
where the matrix elements $Y$ are diagonalizing the neutral Higgs mass$^2$ matrix 
as follows 
$Y M^2_{Higgs}Y^T=diag(m^2_{H^0_1},m^2_{H^0_2},m^2_{H^0_3})$.

For loop of Fig. 1(i), part(c), we find
\beqn
\delta\theta^{(7)}_{k{\ell}}=\frac{g^3m_Z \cos\beta}{4\sqrt 2 \cos\theta_W}\sum_{i=1}^4\sum_{n=1}^3\sum_{m=1}^3
\{Q^{'*}_{i{\ell}}(Y_{n1}-iY_{n3}\sin\beta)-S^{'*}_{i{\ell}}(Y_{n2}-iY_{n3}\cos\beta)\}\nonumber\\
\{Q^{'*}_{ki}(Y_{m1}-iY_{m3}\sin\beta)-S^{'*}_{ki}(Y_{m2}-iY_{m3}\cos\beta)\}\nonumber\\
\{(Y_{n1}+iY_{n3}\sin\beta)(3Y_{m1}-iY_{m3}\sin\beta-4Y_{m2}\tan\beta)\nonumber\\
-2(Y_{m2}-iY_{m3}\cos\beta)(Y_{n2}+iY_{n3}\cos\beta)\}
\frac{m_{\chi^0_i}}{16\pi^2}f(m^2_{\chi^0_i}, m^2_{H^0_m}, m^2_{H^0_n})\nonumber\\
\Delta\tau^{(7)}_{k{\ell}}=\frac{g^3m_Z \cos\beta}{4\sqrt 2 \cos\theta_W}\sum_{i=1}^4\sum_{n=1}^3\sum_{m=1}^3
\{Q^{'}_{{\ell}i}(Y_{n1}+iY_{n3}\sin\beta)-S^{'}_{{\ell}i}(Y_{n2}+iY_{n3}\cos\beta)\}\nonumber\\
\{Q^{'}_{ik}(Y_{m1}+iY_{m3}\sin\beta)-S^{'}_{ik}(Y_{m2}+iY_{m3}\cos\beta)\}\nonumber\\
\{(Y_{n1}+iY_{n3}\sin\beta)(3Y_{m1}-iY_{m3}\sin\beta-4Y_{m2}\tan\beta)\nonumber\\
-2(Y_{m2}-iY_{m3}\cos\beta)(Y_{n2}+iY_{n3}\cos\beta)\}
\frac{m_{\chi^0_i}}{16\pi^2}f(m^2_{\chi^0_i}, m^2_{H^0_m}, m^2_{H^0_n})
\eeqn

For loop of Fig. 1(i), part(d), we find
\beqn
\delta\theta^{(8)}_{k{\ell}}=-\frac{2g^3m_Z\cos\beta}{\sqrt{2}\cos^3\theta_W}\sum_{i=1}^4
R^{'''}_{ki}L^{'''}_{i{\ell}}\frac{m_{\chi^0_i}}{16\pi^2}
f(m^2_{\chi^0_i}, m^2_Z, m^2_Z)\nonumber\\
\Delta\tau^{(8)}_{k{\ell}}=-\frac{2g^3m_Z\cos\beta}{\sqrt{2}\cos^3\theta_W}\sum_{i=1}^4
L^{'''}_{ki}R^{'''}_{i{\ell}}\frac{m_{\chi^0_i}}{16\pi^2}
f(m^2_{\chi^0_i}, m^2_Z, m^2_Z)
\eeqn

For loop of Fig. 1(ii), part(e), we find
\beqn
\delta\theta^{(9)}_{k{\ell}}=-\sum_{i=1}^2\sum_{j=1}^2
\epsilon_{kj}\epsilon^{'*}_{{\ell}i}\phi^{*}_{ij}\cos\beta\sin\beta
\nonumber\\
\frac{m_{\chi^+_i}m_{\chi^+_j}}{16\pi^2}f(m^2_{\chi^+_i}, m^2_{\chi^+_j}, m^2_{H^-})
\nonumber\\
~\Delta\tau^{(9)}_{k{\ell}}=0
\eeqn
The parameters $\epsilon_{ij}$, $\epsilon^{'}_{ij}$ and $\phi_{ij}$ are defined by
\beqn
\epsilon_{ij}=-gX_{4i}V_{j1}^{*}-\frac{g}{\sqrt 2}X_{2i}V_{j2}^{*}-\frac{g}{\sqrt 2} \tan\theta_W X_{1i}V_{j2}^{*}\nonumber\\
~\epsilon^{'}_{ij}=-gX_{3i}^{*}U_{j1}+\frac{g}{\sqrt 2}X_{2i}^{*}U_{j2} +\frac{g}{\sqrt 2} \tan\theta_W X_{1i}^{*}U_{j2}\nonumber\\
\phi_{ij}=-g U_{j2} V_{i1}
\eeqn
where the diagonalizing matrices $U$ and $V$ of the chargino mass matrix
 are defined by
\beq
U^{*} M_{\chi^{+}}V^{-1}=diag(m_{\chi^{+}_1},m_{\chi^{+}_2})
\eeq

For loop of Fig. 1(i), part(e), we find
\beqn
\delta\theta^{(10)}_{k{\ell}}= 
\frac{gm_W}{2\sqrt{2}}\sum_{i=1}^2\epsilon_{ki}\epsilon^{'*}_{{\ell}i}
\cos^2\beta\sin\beta(1+2\sin^2\beta-\cos2\beta\tan^2\theta_W)\nonumber\\
\times\frac{m_{\chi^+_i}}{16\pi^2}f(m^2_{\chi^+_i}, m^2_{H^-}, m^2_{H^-})\nonumber\\
\Delta\tau^{(10)}_{k{\ell}}=
\frac{gm_W}{2\sqrt{2}}\sum_{i=1}^2\epsilon^{'}_{ki}\epsilon^{*}_{{\ell}i}
\cos^2\beta\sin\beta(1+2\sin^2\beta-\cos2\beta\tan^2\theta_W)\nonumber\\
\times\frac{m_{\chi^+_i}}{16\pi^2}f(m^2_{\chi^+_i}, m^2_{H^-}, m^2_{H^-})
\eeqn

For loop of Fig. 1(i), part(f), we find
\beqn
\delta\theta^{(11)}_{k{\ell}}=
-\sum_{i=1}^2\frac{g^3}{\sqrt{2}}m_W\cos\beta R_{ki}L^{*}_{{\ell}i}\frac{m_{\chi^+_i}}{4\pi^2}f(m^2_{\chi^+_i}, m^2_{W^-}, m^2_{W^-})\nonumber\\
\Delta\tau^{(11)}_{k{\ell}}=
-\sum_{i=1}^2\frac{g^3}{\sqrt{2}}m_W\cos\beta R^{*}_{{\ell}i}L_{ki}\frac{m_{\chi^+_i}}{4\pi^2}f(m^2_{\chi^+_i}, m^2_{W^-}, m^2_{W^-})
\eeqn
where $L$ and$R$ are defined as
\beqn
L_{ij}=-\frac{1}{\sqrt 2}X^{*}_{4i} V^{*}_{j2} +X^{*}_{2i} V^{*}_{j1}\nonumber\\
~R_{ij}=\frac{1}{\sqrt 2}X_{3i} U_{j2} +X_{2i} U_{j1}
\eeqn

For loop of Fig. 1(ii), part(f), we find
\beqn
\delta\theta^{(12)}_{k{\ell}}=0\nonumber\\
\Delta\tau^{(12)}_{k{\ell}}=
\sum_{i=1}^2\sum_{j=1}^2 g^2 \phi^{*}_{ij}L_{kj}R^{*}_{{\ell}i}
\frac{m_{\chi^+_i}m_{\chi^+_j}}{4\pi^2}
f(m^2_{\chi^+_i}, m^2_{\chi^+_j}, m^2_{W^-})
\eeqn

For loop of Fig. 1(ii), part(g), we find
\beqn
\delta\theta^{(13)}_{k{\ell}}=0\nonumber\\
\Delta\tau^{(13)}_{k{\ell}}=\sum_{i=1}^2\frac{h_{\tau}m^2_{\tau}}{8\pi^2}
(\beta_{\tau{\ell}}D_{\tau1i}+\alpha^{*}_{\tau{\ell}}D_{\tau2i})\nonumber\\
\times (\alpha^{*}_{\tau k}D^{*}_{\tau1i}-\gamma^{*}_{\tau k}D^*_{\tau2i}) f(m^2_{\tilde{\tau}_i}, m^2_{\tau}, m^2_{\tau})
\eeqn
where 
\beq
h_{\tau}=\frac{gm_{\tau}}{\sqrt 2 m_W \cos\beta}
\eeq

For loop of Fig. 1(i), part(g), we find
\beqn
\delta\theta^{(14)}_{k{\ell}}=-\sum_{i=1}^2\sum_{j=1}^2 
\frac{m_{\tau}}{8\pi^2}
H^{*}_{{\tau}ji} (\alpha_{\tau{\ell}}D_{\tau1i}-\gamma_{\tau{\ell}}D_{\tau2i})\nonumber\\
\times (\beta^{*}_{\tau k}D^{*}_{\tau1i}+\alpha_{\tau k}D^*_{\tau2j}) f(m^2_{\tau}, m^2_{\tilde{\tau}_i}, m^2_{\tilde{\tau}_j})\nonumber\\
\Delta\tau^{(14)}_{k{\ell}}=-\sum_{i=1}^2\sum_{j=1}^2 
\frac{m_{\tau}}{8\pi^2}
H^{*}_{{\tau}ji} (\beta_{\tau{\ell}}D_{\tau1i}+\alpha^{*}_{\tau{\ell}}D_{\tau2i})\nonumber\\
\times (\alpha^{*}_{\tau k}D^{*}_{\tau1j}-\gamma^{*}_{\tau k}D^*_{\tau2j}) f(m^2_{\tau}, m^2_{\tilde{\tau}_i}, m^2_{\tilde{\tau}_j})
\eeqn

and $H_{{\tau}ij}$ is given by
\beqn
H_{{\tau}ij}=-\frac{gM_Z}{\sqrt 2 \cos\theta_W}
((-\frac{1}{2}+\sin^2\theta_W)D_{\tau1i}^{*}D_{\tau1j}
-\sin^2\theta_W D_{\tau2i}^{*}D_{\tau2j})\cos\beta \nonumber\\
-\frac{gm^2_{\tau}}{\sqrt 2 m_W \cos\beta}(D_{\tau1i}^{*}D_{\tau1j}+
D_{\tau2i}^{*}D_{\tau2j})-\frac{g m_{\tau} A_{\tau}}{\sqrt 2 m_W \cos\beta}
D_{\tau2i}^{*}D_{\tau1j}
\eeqn

The loop corrections for $\delta\theta_{k{\ell}}$ and 
$\Delta\tau_{k{\ell}}$ are given by
\beqn
\delta\theta_{k{\ell}}=\sum_{n=1}^{14}\delta\theta^{(n)}_{k{\ell}} \nonumber\\
~\Delta\tau_{k{\ell}}=\sum_{n=1}^{14}\Delta\tau^{(n)}_{k{\ell}}
\eeqn

\subsection{Loop analysis of $\Delta \theta_{k{\ell}}$ and  $\delta \tau_{k{\ell}}$}

We do the same analysis of Figure 2 as for Figure 1. We write down here the final results for both corrections from the fourteen loops together. The corrections are written in the same order of the loops in Figure 2.
\beqn
\Delta \theta_{k{\ell}}=-\sum_{i=1}^2\sum_{j=1}^2 
\frac{m_t}{8\pi^2}
E_{ji} (\alpha_{t{\ell}}D_{t1j}-\gamma_{t{\ell}}D_{t2j})
(\beta^{*}_{tk}D^{*}_{t1i}+\alpha_{tk}D^*_{t2i}) f(m^2_t, m^2_{\tilde{t}_i}, m^2_{\tilde{t}_j})\nonumber\\
-\sum_{i=1}^2\sum_{j=1}^2 
\frac{m_b}{8\pi^2}
G_{ji} (\alpha_{b{\ell}}D_{b1j}-\gamma_{b{\ell}}D_{b2j})(\beta^{*}_{bk}D^{*}_{b1i}+\alpha_{bk}D^*_{b2i}) f(m^2_b, m^2_{\tilde{b}_i}, m^2_{\tilde{b}_j})\nonumber\\
+\sum_{j=1}^2 \frac{m^2_t h_t}{8\pi^2}(\alpha_{t{\ell}}D_{t1j}-\gamma_{t{\ell}}D_{t2j})(\beta^{*}_{tk}D^{*}_{t1j}+\alpha_{t k}D^*_{t2j}) f(m^2_{\tilde{t}_j}, m^2_{t}, m^2_{t})\nonumber\\
+0\nonumber\\
+0\nonumber\\
+0\nonumber\\
+\frac{g^3m_Z \cos\beta}{4\sqrt 2 \cos\theta_W}\sum_{i=1}^4\sum_{n=1}^3\sum_{m=1}^3
\{Q^{'*}_{i{\ell}}(Y_{n1}-iY_{n3}\sin\beta)-S^{'*}_{i{\ell}}(Y_{n2}-iY_{n3}\cos\beta)\}\nonumber\\
\{Q^{'*}_{ki}(Y_{m1}-iY_{m3}\sin\beta)-S^{'*}_{ki}(Y_{m2}-iY_{m3}\cos\beta)\}
\{\tan\beta(Y_{n2}-iY_{n3}\cos\beta)(3Y_{m2}+iY_{m3}\cos\beta)\nonumber\\
-4Y_{n1}(Y_{m2}-iY_{m3}\cos\beta)-2\tan\beta(Y_{m1}-iY_{m3}\sin\beta)
(Y_{n1}+iY_{n3}\sin\beta)\}
\frac{m_{\chi^0_i}}{16\pi^2}f(m^2_{\chi^0_i}, m^2_{H^0_m}, m^2_{H^0_n})\nonumber\\
-\frac{2g^3m_Z\sin\beta}{\sqrt{2}\cos^3\theta_W}\sum_{i=1}^4
R^{'''}_{ki}L^{'''}_{i{\ell}}\frac{m_{\chi^0_i}}{16\pi^2}
f(m^2_{\chi^0_i}, m^2_Z, m^2_Z)\nonumber\\
+0\nonumber\\
+\frac{gm_W}{2\sqrt{2}}\sum_{i=1}^2\epsilon_{ki}\epsilon^{'*}_{{\ell}i}
\cos\beta\sin^2\beta(1+2\cos^2\beta+\cos2\beta\tan^2\theta_W)\frac{m_{\chi^+_i}}{16\pi^2}f(m^2_{\chi^+_i}, m^2_{H^-}, m^2_{H^-})\nonumber\\
-\sum_{i=1}^2\frac{g^3}{\sqrt{2}}m_W\sin\beta R_{ki}L^{*}_{{\ell}i}\frac{m_{\chi^+_i}}{4\pi^2}f(m^2_{\chi^+_i}, m^2_{W^-}, m^2_{W^-})\nonumber\\
+\sum_{i=1}^2\sum_{j=1}^2 g^2 \psi_{ij}R_{kj}L^{*}_{{\ell}i}
\frac{m_{\chi^+_i}m_{\chi^+_j}}{4\pi^2}
f(m^2_{\chi^+_i}, m^2_{\chi^+_j}, m^2_{W^-})\nonumber\\
+0\nonumber\\
-\sum_{i=1}^2\sum_{j=1}^2 
\frac{m_{\tau}}{8\pi^2}
G_{{\tau}ij} (\alpha_{\tau{\ell}}D_{\tau1i}-\gamma_{\tau{\ell}}D_{\tau2i})(\beta^{*}_{\tau k}D^{*}_{\tau1i}+\alpha_{\tau k}D^*_{\tau2j}) f(m^2_{\tau}, m^2_{\tilde{\tau}_i}, m^2_{\tilde{\tau}_j})
\eeqn

The corrections $\delta \tau_{k{\ell}}$ are given by
\beqn
\delta \tau_{k{\ell}}=-\sum_{i=1}^2\sum_{j=1}^2 
\frac{m_t}{8\pi^2}
E_{ji} (\beta_{t{\ell}}D_{t1j}+\alpha^{*}_{t{\ell}}D_{t2j})(\alpha^{*}_{tk}D^{*}_{t1i}-\gamma^{*}_{tk}D^*_{t2i}) f(m^2_t, m^2_{\tilde{t}_i}, m^2_{\tilde{t}_j})\nonumber\\
-\sum_{i=1}^2\sum_{j=1}^2 
\frac{m_b}{8\pi^2}
G_{ji} (\beta_{b{\ell}}D_{b1j}+\alpha^{*}_{b{\ell}}D_{b2j})(\alpha^{*}_{bk}D^{*}_{b1i}-\gamma^{*}_{bk}D^*_{b2i}) f(m^2_b, m^2_{\tilde{b}_i}, m^2_{\tilde{b}_j})\nonumber\\
+0\nonumber\\
+0\nonumber\\
+\sum_{i=1}^4\sum_{j=1}^4\frac{g^3}{2\pi^2\cos^2\theta_W}
S^{'}_{ij}L^{'''}_{kj}R^{'''}_{i{\ell}}m_{\chi^0_i}m_{\chi^0_j}f(m^2_{\chi^0_i}, m^2_{\chi^0_j}, m^2_{Z})\nonumber\\
-\sum_{i=1}^4\sum_{j=1}^4\sum_{n=1}^3\frac{g^3}{4\pi^2}
S^{'}_{ij}
\{Q^{'}_{{\ell}i}(Y_{n1}+iY_{n3}\sin\beta)-S^{'}_{{\ell}i}(Y_{n2}+iY_{n3}\cos\beta)\}\nonumber\\
\{Q^{'}_{jk}(Y_{n1}+iY_{n3}\sin\beta)-S^{'}_{jk}(Y_{n2}+iY_{n3}\cos\beta)\}
m_{\chi^0_i}m_{\chi^0_j}f(m^2_{\chi^0_i}, m^2_{\chi^0_j}, m^2_{H^0_n})\nonumber\\
+\frac{g^3m_Z \cos\beta}{4\sqrt 2 \cos\theta_W}\sum_{i=1}^4\sum_{n=1}^3\sum_{m=1}^3
\{Q^{'}_{{\ell}i}(Y_{n1}+iY_{n3}\sin\beta)-S^{'}_{{\ell}i}(Y_{n2}+iY_{n3}\cos\beta)\}\nonumber\\
\{Q^{'}_{ik}(Y_{m1}+iY_{m3}\sin\beta)-S^{'}_{ik}(Y_{m2}+iY_{m3}\cos\beta)\}
\{\tan\beta(Y_{n2}-iY_{n3}\cos\beta)(3Y_{m2}+iY_{m3}\cos\beta)\nonumber\\
-4Y_{n1}(Y_{m2}-iY_{m3}\cos\beta)-2\tan\beta (Y_{m1}-iY_{m3}\sin\beta)
(Y_{n1}+iY_{n3}\sin\beta)\}
\frac{m_{\chi^0_i}}{16\pi^2}f(m^2_{\chi^0_i}, m^2_{H^0_m}, m^2_{H^0_n})\nonumber\\
-\frac{2g^3m_Z\sin\beta}{\sqrt{2}\cos^3\theta_W}\sum_{i=1}^4
L^{'''}_{ki}R^{'''}_{i{\ell}}\frac{m_{\chi^0_i}}{16\pi^2}
f(m^2_{\chi^0_i}, m^2_Z, m^2_Z)\nonumber\\
-\sum_{i=1}^2\sum_{j=1}^2
\epsilon^{'}_{kj}\epsilon^{*}_{{\ell}i}\psi_{ij}\cos\beta\sin\beta
\frac{m_{\chi^+_i}m_{\chi^+_j}}{16\pi^2}f(m^2_{\chi^+_i}, m^2_{\chi^+_j}, m^2_{H^-})
\nonumber\\
+\frac{gm_W}{2\sqrt{2}}\sum_{i=1}^2\epsilon^{'}_{ki}\epsilon^{*}_{{\ell}i}
\cos\beta\sin^2\beta(1+2\cos^2\beta+\cos2\beta\tan^2\theta_W)\frac{m_{\chi^+_i}}{16\pi^2}f(m^2_{\chi^+_i}, m^2_{H^-}, m^2_{H^-})\nonumber\\
-\sum_{i=1}^2\frac{g^3}{\sqrt{2}}m_W\sin\beta R^{*}_{{\ell}i}L_{ki}\frac{m_{\chi^+_i}}{4\pi^2}f(m^2_{\chi^+_i}, m^2_{W^-}, m^2_{W^-})\nonumber\\
+0\nonumber\\
+0\nonumber\\
-\sum_{i=1}^2\sum_{j=1}^2 
\frac{m_{\tau}}{8\pi^2}
G_{{\tau}ij} (\beta_{\tau{\ell}}D_{\tau1i}+\alpha^{*}_{\tau{\ell}}D_{\tau2i})(\alpha^{*}_{\tau k}D^{*}_{\tau1j}-\gamma^{*}_{\tau k}D^*_{\tau2j}) f(m^2_{\tau}, m^2_{\tilde{\tau}_i}, m^2_{\tilde{\tau}_j})
\eeqn

where $G_{ij}$, $E_{ij}$, $h_t$, $\psi_{ij}$ and $G_{\tau ij}$ are given by
\beqn
G_{ij}=\frac{gM_Z}{\sqrt 2 \cos\theta_W}
((-\frac{1}{2}+\frac{1}{3}\sin^2\theta_W)D_{b1i}^{*}D_{b1j}
-\frac{1}{3}\sin^2\theta_W D_{b2i}^{*}D_{b2j})\sin\beta \nonumber\\
+\frac{g m_b \mu}{\sqrt 2 m_W \cos\beta}
D_{b1i}^{*}D_{b2j}\nonumber\\
E_{ij}=\frac{gM_Z}{\sqrt 2 \cos\theta_W}
((\frac{1}{2}-\frac{2}{3}\sin^2\theta_W)D_{t1i}^{*}D_{t1j}
+\frac{2}{3}\sin^2\theta_W D_{t2i}^{*}D_{t2j})\sin\beta \nonumber\\
-\frac{gm^2_t}{\sqrt 2 m_W \sin\beta}(D_{t1i}^{*}D_{t1j}+
D_{t2i}^{*}D_{t2j})-\frac{g m_t A_t}{\sqrt 2 m_W \sin\beta}
D_{t2i}^{*}D_{t2j}\nonumber\\
h_{t}=\frac{gm_t}{\sqrt 2 m_W \sin\beta},
\psi_{jk}=-g U_{k1} V_{j2}\\
G_{\tau ij}=\frac{gM_Z}{\sqrt 2 \cos\theta_W}
((-\frac{1}{2}+\sin^2\theta_W)D_{\tau1i}^{*}D_{\tau1j}
-\sin^2\theta_W D_{\tau2i}^{*}D_{\tau2j})\sin\beta \nonumber\\
+\frac{g m_{\tau} \mu}{\sqrt 2 m_W \cos\beta}
D_{\tau1i}^{*}D_{\tau2j}
\eeqn

\section{Neutral Higgs decays including loop effects}

We summarize now the result of the analysis. 
Thus ${\cal{L}}_{eff}$ of Eq.~({\ref{t1}}) may be written as follows
\begin{equation}
{\cal {L}}_{eff}=H^{0}_{m}\overline{\chi^0_k}(\alpha^{mS}_{k{\ell}}+\gamma_{5}\alpha^{mP}_{k{\ell}})\chi^0_{\ell}+ 
H.c
\end{equation}
where
\begin{equation}
\alpha^{mS}_{k{\ell}}=\frac{1}{2\sqrt 2}\{(Y_{m1}-iY_{m3}\sin\beta)(\theta_{k{\ell}}+\delta \theta_{k{\ell}}+\Delta \tau_{k{\ell}})+
(Y_{m2}+iY_{m3}\cos\beta)(\tau_{k{\ell}}+\Delta \theta_{k{\ell}}+\delta \tau_{k{\ell}})\}
\label{scoupling}
\end{equation}
and where
\begin{equation}
\alpha^{mP}_{k{\ell}}=\frac{1}{2\sqrt 2}\{(Y_{m2}+iY_{m3}\cos\beta)(\tau_{k{\ell}}+\delta \tau_{k{\ell}}-\Delta \theta_{k{\ell}})+
(Y_{m1}-iY_{m3}\sin\beta)(-\theta_{k{\ell}}+\Delta \tau_{k{\ell}}-\delta \theta_{k{\ell}})\}
\label{pcoupling}
\end{equation}
Next we discuss the implications of the above result for the 
decay of the neutral Higgs.

The partial width of the decay $H^{0}_m\rightarrow\chi^0_k\chi^0_{\ell}$ is given by
\beqn
\Gamma_{mk{\ell}}(H^{0}_m\rightarrow\chi^0_k\chi^0_{\ell})=\frac{1}{\pi 
M^3_{H^0_m}(1+\delta_{k\ell})}
\sqrt{[(m^2_{\chi^0_{\ell}}+m^2_{\chi^{0}_k}-M^2_{H^{0}_m})^2
-4m^2_{\chi^{0}_k}m^2_{\chi^0_{\ell}}]}\nonumber\\
\{\frac{1}{2}(|\alpha^{mS}_{k{\ell}}|^2+|\alpha^{mP}_{k{\ell}}|^2)
(M^2_{H^{0}_m}-m^2_{\chi^{0}_k}-m^2_{\chi^{0}_{\ell}})
-\frac{1}{2}(|\alpha^{mS}_{k{\ell}}|^2-|\alpha^{mP}_{k{\ell}}|^2)
(2m_{\chi^{0}_k}m_{\chi^{0}_{\ell}})\}
\label{branching}
\eeqn
The neutral Higgs bosons can decay into different modes. However, there are
 important channels for this decay to occur,
$\bar{b} b$, $\bar{t} t$, $\bar{s} s$, $\bar{c} c$, $\bar{\tau} \tau$, 
 $\chi^{+}_i \chi^{-}_j$ and $\chi^0_i \chi^0_j$.
The other channels of neutral Higgs decay are the decaying
 modes into the other fermions of the SM,  squarks, sleptons,
other Higgs bosons, W and Z boson pairs, one Higgs and a vector boson, 
$\gamma \gamma$ pairs and finally into the gluonic decay i.e, $H^0_m\rightarrow g g$.
The lightest SM fermions channels could be ignored for the smallness 
of their couplings.
We choose the region in the parameter space where we can ignore
the other channels which either are not allowed kinematically or suppressed by 
their couplings. Thus in this work, squarks and sleptons are too heavy to be
relevant in neutral Higgs decay. The neutral Higgs decays into
non-supersymmetric final states that involve gauge bosons and/or 
other Higgs bosons are ignored as well.
In the region of large $\tan\beta$, 
these decays are very small and can be neglected as final states \cite{gunion}.

We calculate the radiative corrected partial decay widths of the important channels
mentioned above. In the case of CP violating case under investigation we 
use  the 
analysis of \cite{me1},
for the radiatively corrected $\Gamma$ of neutral Higgs into quarks, leptons and chargino pairs.  For the radiatively  corrected decay width  into neutralino we use the current analysis.
We define 
\beq
\Delta\Gamma_{mk{\ell}}=\frac{\Gamma(H^0_m\rightarrow\chi_k^0\chi_{\ell}^0)-\Gamma^0(H^0_m\rightarrow\chi_k^0\chi_{\ell}^0)}{\Gamma^0(H^0_m\rightarrow\chi_k^0\chi_{\ell}^0)}
\label{gammad}
\eeq
where the first term in the numerator is the decay width including
the full loop corrections and the second term is the decay width
evaluated at the tree level.
Finally to investigate the size of the loop effects on 
the  branching  ratios of the neutral Higgs decay we define 
the following quantity
\beq
\Delta Br_{mk{\ell}}=\frac{Br(H^0_m\rightarrow\chi_k^0\chi_{\ell}^0)-Br^0(H^0_m\rightarrow\chi_k^0\chi_{\ell}^0)}{Br^0(H^0_m\rightarrow\chi_k^0\chi_{\ell}^0)}
\label{bran1}
\eeq
where the first term in the numerator is the branching ratio including
the full loop corrections and the second term is the branching ratio
evaluated at the tree level.
The analysis of this section is utilized in Sec.(4) where we 
give a numerical analysis of the size of the loop effects and
discuss the effect of the loop corrections on the branching 
ratios.
\section{NUMERICAL ANALYSIS}

In this section we investigate the size of the loop corrections
on the partial decay widths and the branching ratios of the neutral
Higgs bosons decay into neutralinos.
The analysis of Sec. 2 and Sec. 3 is quite general and valid for the minimal 
supersymmetric standard model. For the sake of numerical analysis we will limit 
the parameter space by working within the framework of the SUGRA model 
\cite{sugra1}. Specifically we will work within the framework of the the extended
nonuniversal mSUGRA model including CP phases. 
We take as our parameter space at the grand unification scale to be 
the following: the universal scalar mass $m_0$, the universal gaugino mass $m_{1/2}$, 
the universal trilinear coupling 
$|A_0|$, the ratio of the Higgs vacuum expectation values $\tan\beta=<H_2>/<H_1>$
where $H_2$ gives mass to the up quarks and $H_1$ gives mass to the down 
quarks and the leptons.
In addition, we take for CP phases the following: the phase $\theta_{\mu}$ of the Higgs mixing parameter $\mu$, the phase $\alpha_{A_0}$ of the trilinear coupling $A_0$ and the phases $\xi_i(i=1,2,3)$ of the $SU(3)_C$, $SU(2)_L$ and $U(1)_Y$ gaugino masses.
In this analysis the electroweak symmetry is broken
by radiative effects which allows one to determine the
magnitude of $\mu$ by fixing $M_Z$. In the analysis we use
one loop renormalization group (RGEs) equations for the evolution
of the soft SUSY breaking parameters and for the parameter $\mu$, and two loop RGEs for the gauge and Yukawa couplings.
In the numerical analysis we compute the loop corrections and also analyze 
their dependence on the phases. The masses of particles involved in the analysis
are ordered as follows: 
for neutralinos $m_{\chi^0_1}<m_{\chi^0_2}<m_{\chi^0_3}<m_{\chi^0_4}$ and
for the neutral Higgs $(m_{H_1},m_{H_2},m_{H_3})\rightarrow (m_H,m_h,m_A)$ in the limit of no CP mixing where $m_H$ is the heavy CP even Higgs,
$m_h$ is the light CP even Higgs, and $m_A$ is the CP odd Higgs.

We first discuss the size of the loop corrections of the partial decay width defined
in Eq.(\ref{gammad}). As was mentioned before, the loop corrected partial widths of 
the neutral Higgs decay into neutralinos have been investigated in the absence
of CP violating phases \cite{eberl,ren}. The magnitude of the corrections 
in these analyses is of the order of $\sim10\%$ of the tree level value. 
The current analysis supports this result. In 
Fig. (\ref{epsfig7a})
, we give a plot of
$\Delta\Gamma_{113}$
as functions of $\tan\beta$ for the specific set of inputs given in the figure caption. We notice that the partial decay width gets a change of 
$2\sim12\%$ of its tree level value.
The role played by $\tan\beta$ in this analysis is complicated and is
coming from different regions in the analysis. First of all, it affects the
spectrum and couplings of neutral Higgs with neutralinos at tree level through
the diagonalizing matrices of both neutral Higgs bosons and neutralino.
We also find that $\tan\beta$ is playing a crucial rule at the one loop level
analysis. The neutral Higgs mass$^2$ matrix receives corrections from the stop, 
sbottom, chargino and neutralino sectors and these corrections are sensitive to
the value of $\tan\beta$. We also see the explicit and implicit  
effects of $\tan\beta$ in
the loop corrected couplings of 
neutralinos with neutral Higgs presented 
in Eq. (\ref{scoupling}) and Eq. (\ref{pcoupling}) for $\alpha^{mS}_{k{\ell}}$
and $\alpha^{mP}_{k{\ell}}$ respectively.
 We also notice that the CP violating 
phase $\theta_{\mu}$ can affect the value of this change. This effect has not 
been discussed in the previous analyses because these analyses have been
carried out for the CP conservation case.
We can also trace down the role played by the phase $\theta_{\mu}$ in the
analysis.
We can see that, $\theta_{\mu}$ affects the tree level of analysis through
its presence in the neutralino mass matrix and at loop level where it can produce
mixing in the neutral Higgs sector and also affects the radiative corrected
couplings between the neutralinos and neutral Higgs bosons.
 In the limit where CP violating phases
are set to zero and by using the same inputs of \cite{eberl}, we 
were able to have a fair agreement with with their Figs. 2-4, 6. In the work
of \cite{ren} only 8 out of 28 diagrams of the current analysis are calculated.
By including these diagrams only in the comparison, our analysis
is in fair agreement with their Figs.2, 3, 5, 7 and 9 for their inputs.

Now we compute the loop correction effects of the branching ratios 
of the neutral Higgs decays into neutralinos.
The branching ratio of a decay mode is the ratio between the partial decay
rate of this mode and the total decay rate for all possible channels.
In the parameter space we are investigating, these channels are decays into
charginos, heavy quarks, taus, and neutralinos. 
In Figs. (\ref{epsfig7b}) and (\ref{epsfig7c}) we give a plot of 
$\Delta Br_1\to \chi_2^0\chi_2^0$  and $\Delta Br_3\to \chi_1^0\chi_3^0$ as functions of
$m_{1/2}$ for the specific set of inputs given in the captions of these figures. 
We first notice that the loop correction of the branching
ratios can reach as high as $35\%$  of the tree level value for the case of $H_1$ boson
and as high as $55\%$ for the case of $H_3$ boson.
We also can see the effect of the CP violating phase
$\theta_{\mu}$ in these
two figures. In the branching ratio study, this CP violating
phase can affect many decay modes of neutral Higgs into
different quarks and leptons via  radiative corrections of these
modes. It can affect both tree and loop level of the analysis
in the cases of decays into charginos and neutralinos due to the presence of the parameter $\mu$ in the chargino, neutralino and sfermion mass matrices.
The role played by the parameter $m_{1/2}$ is mainly through
the chargino and neutralino mass matrices since the gaugino
masses $\tilde{m}_1$ and $\tilde{m}_2$ are originating from
$m_{1/2}$ at GUT scale. The parameter $m_{1/2}$ is also
affecting the evolution of the other soft supersymmetry breaking parameters like the trilinear couplings $A_f$ from
GUT scale down to the electroweak scale. 

In Figs. (\ref{epsfig7d}) and (\ref{epsfig7e}) we give a plot of 
$\Delta Br_1\to \chi_1^0\chi_2^0$  and $\Delta Br_3\to \chi_2^0\chi_2^0$ as functions of
$\theta_{\mu}$ for the specific set of inputs given in the captions of these figures. 
We notice in these two figures that the loop corrections of the branching
ratios for these modes can reach as high as $35\%$  of the tree level value. 
We see here again the effect of the CP violating phase $\theta_{\mu}$ on the corrections of branching ratio for these
decay modes. In the case of $H_3$ decay, one can see that $\theta_{\mu}$ affects not only the magnitude of 
$\Delta Br_3\to \chi_2^0\chi_2^0$ but also its sign 
depending on $\theta_{\mu}$. The analysis of these two figures 
also shows  the importance of the parameter $\tan\beta$ in
the loop corrections for these the branching ratios. 
This parameter is important at tree level through 
neutral Higgs couplings with different quarks and leptons and through
the diagonalization of the neutral Higgs, chargino and neutralino mass matrices. 
At one loop level, it affects both neutral Higgs 
spectrum and couplings  with different fields.

In Figs. (\ref{epsfig7f}) and (\ref{epsfig7g}) we give a plot of 
$\Delta Br_1\to \chi_1^0\chi_3^0$  and $\Delta Br_3\to \chi_1^0\chi_2^0$ as functions of
$\alpha_{0}$ for the specific set of inputs given in the captions of these figures. 
We notice in these two figures that the loop correction of the
branching ratios for these modes can reach as high as $40\%$ of the tree level.
The effects of the magnitude of $|A_0|$ and its CP violating phase 
 are clear
in both modes and could be 
understood form the effect of the trilinear couplings on
the squark and slepton mass$^2$ matrices in the stop case through $A_t$,
 in the sbottom
case through $A_b$, in the stau case through the parameter $A_{\tau}$.

In Figs. (\ref{epsfig7h}) and (\ref{epsfig7i}) we give a plot of 
$\Delta Br_1\to \chi_1^0\chi_3^0$  and $\Delta Br_3\to \chi_1^0\chi_2^0$ as functions of
$\xi_2$ for the specific set of inputs given in the captions of these figures. 
Here we find that $\xi_2$ phase has a smaller effect on the loop corrections.
The reason for this could be understood qualitatively from the fact
that the chargino and neutralino loops that carry the effect of this phase
are correcting the tree level of the analysis less than that of the other
loops in this region of the parameter space.

\section{RELEVANCE OF RESULTS AT LHC}

The production of the MSSM Higgs particles at the Large Hadron Collider LHC ($\sqrt{s}=14$ TeV)
occurs via gluon fusion $gg\rightarrow H_i$ and the associated production mechanism
$gg+q\bar{q}\rightarrow b\bar{b} H_i$. The cross section of these processes can reach few
tens of pb at large $\tan\beta$ region and for a 
moderate Higgs masses $\sim 500$ GeV. For integrated luminosity (10) 100 $fb^{-1}$ in the
(low) high luminosity option, $\sigma=1$ pb would correspond to ($10^4$) $10^5$ events \cite{djouadi}.
These Higgs particles once produced, can decay into many channels and one of them is the
channel considered here, the neutralino one.

The decay of the heavy MSSM Higgs bosons to 
neutralinos
could be observed at LHC. When 
$\chi^0$ decay channels are open, their branching ratios can be
close to $\sim 20\%$ \cite{djouadi1} and that gives an opportunity for experimental
analysis of the MSSM parameter space. The authors of \cite{moortgat}, study
the decays of $H_1$ and $H_3$ after their
production at LHC into two next-to-lightest neutralinos $\chi^0_2$, with
each of the neutralinos in turn decaying to two Standard Model fermions
along with the lightest neutralino 
$\chi^0_1$,
assumed to be the lightest supersymmetric particle (the LSP) and 
carries missing energy.
The two
fermions will most often be quarks, leading to two jets and missing $E_T$ in the final state. To obtain a clean
signature, one should only focus on the case where the two SM fermions are leptons.
Thus the process under consideration is
\beq
H_1, H_3 \rightarrow \chi^0_2\chi^0_2\rightarrow 4\ell^{\pm} +E_T^{miss}~~(\ell=e,\mu)
\eeq
The above process provides a clear signature containing 
 two
pairs of leptons with opposite sign and same flavor, in addition to a substantial amount of
missing energy due to the escaping lightest neutralino.
In their analysis, the authors of \cite{moortgat} show that one can distinguish this signal from the
(mainly SUSY) background for values of $\tan\beta=5-40$. 
Their analysis for the decay of Heavy Higgs bosons into neutralinos is based on
the HDECAY package \cite{hdecay}. This analysis does not take into account the
loop corrections of the neutral Higgs vertices with neutralinos and is carried
out in the CP conserving scenario. They also study the decay of neutralinos
into leptons in the limit of vanishing CP phases.
In the case (2) of the first paper of \cite{moortgat}, the author used the inputs
$M_2=180$, $M_1=100$, $\mu=500$, $m_{\tilde{\ell}}=250$ and $M_{\tilde{q},\tilde{g}}=1000$ GeV.
It is shown in Fig. (6) of \cite{moortgat}, for integrated luminosity of 100 $fb^{-1}$,
that the expectation to discover the Higgs bosons with a clear and
visible signature over the background occurs for $m_A=380$ GeV and $\tan\beta=10$.
Now by putting these parameters by hand in our analysis with setting all the CP phases
to zero, we get for $\Delta Br_{322}$, defined by Eq. (\ref{bran1}), the value of $\sim-25\%$.
 So the
tree value of the branching ratio that was used in the analysis of \cite{moortgat}
would have been suppressed by radiative corrections of the above percentage and that 
would of course change the output of the analysis.

In the analysis of \cite{bisset}, the authors  investigate the same four-lepton signal with missing
energy at LHC. In their top  Fig. 3,  they use for their inputs,  $\tan\beta=20$, 
$M_1=\frac{5}{3}\tan^2\theta_W M_2$, $m_A=400$, $m_{\tilde{\ell}}=150$, $M_{\tilde{q}}=1000$,
$M_{\tilde{g}}= 800$, $A_{\tau}=A_{l}=0$ and $m_{\tilde{\tau}}=250$ GeV. For the parameter point 
$\mu=-200$ GeV and $M_2=200$ GeV, one has 
$\sigma(pp\rightarrow H_1,H_3)\times Br(H_1,H_3\rightarrow 4{\ell}+E_T^{miss})= 37 fb$.  Thus for
 an integrated luminosity of $100 fb^{-1}$, the event number can reach 3700 events before applying selection cuts.
 In this figure and for
this point, the four-lepton signal originates mainly through $\chi^0_2 \chi^0_2$ channel. 
By calculating the corrections to the branching ratios in our analysis for this input 
but with no CP violating phases, 
one finds that the branching ratio corrections $\Delta Br_{322}$ and $\Delta Br_{122}$  are $+28\%$ and $+24\%$ respectively.
 The authors of \cite{bisset} did not take into account the loop corrections
to the  branching ratios of neutral Higgs into the neutralino 
 and thus the inclusion of these corrections in their  analysis would enhance the
event number at LHC.

We note further, that the couplings of the Higgs bosons to the SM particles and their
supersymmetric partners are modified by the CP violation phases. The Higgs
boson masses and their CP properties are modified as well from those predicted
in the CP conserving case. Thus the cross sections for MSSM Higgs particles production
and their decay signatures could also be much more complicated  than in the CP 
preserving 
 scenario. So an analysis that considers the Higgs bosons production 
and their detection in the environment of LHC with CP violating phases
would be much more involved and is beyond the scope of this  paper.

\section{CONCLUSION}
In this paper we have worked out the loop corrections to  $\chi^{0}_k\chi^{0}_{\ell}H^0_m$ couplings within MSSM.
This analysis extends previous analysis of supersymmetric loop corrections to
the couplings of neutral Higgs bosons with charginos and with standard model
fermions within minimal supersymmetric standard models including the full set
of allowed CP phases. The result of the analysis is then applied to the computation
of the decay of the neutral Higgs bosons to neutralino pairs.
In the absence of loop corrections, the lightest Higgs boson mass
is less than $M_Z$ and including these corrections can lift the lightest
Higgs mass above $M_Z$. In the CP invariance scenario the spectrum of the
neutral Higgs sector consists of two CP even Higgs bosons and one CP odd Higgs boson.
 With
the inclusion of CP phases, the Higgs boson mass eigenstates are
no longer CP even and CP odd states when loop corrections to the
Higgs boson mass matrix are included.
Further, inclusion of loop corrections to the couplings of neutralinos
with neutral Higgs is in general dependent on CP phases. Thus
the decays of neutral Higgs into neutralinos can be sensitive
to the loop corrections and to the CP violating phases.
The effect of the supersymmetric loop corrections is found to
to be in the range of $\sim10\%$ for the partial decay 
width. For the branching ratios it is found to be
 be rather large, as much as $50\%$ in some regions of the
parameter space. The effect of CP phases on the modifications
of the partial decay width and the branching ratio is found to be substantial in some
regions of the MSSM parameter space. 
Specific attention is paid to the neutralino decay mode that can lead
to a four-lepton signal.

\begin{figure}[t]
\hspace*{-0.8in}
\centering
\includegraphics[width=18cm,height=18cm]{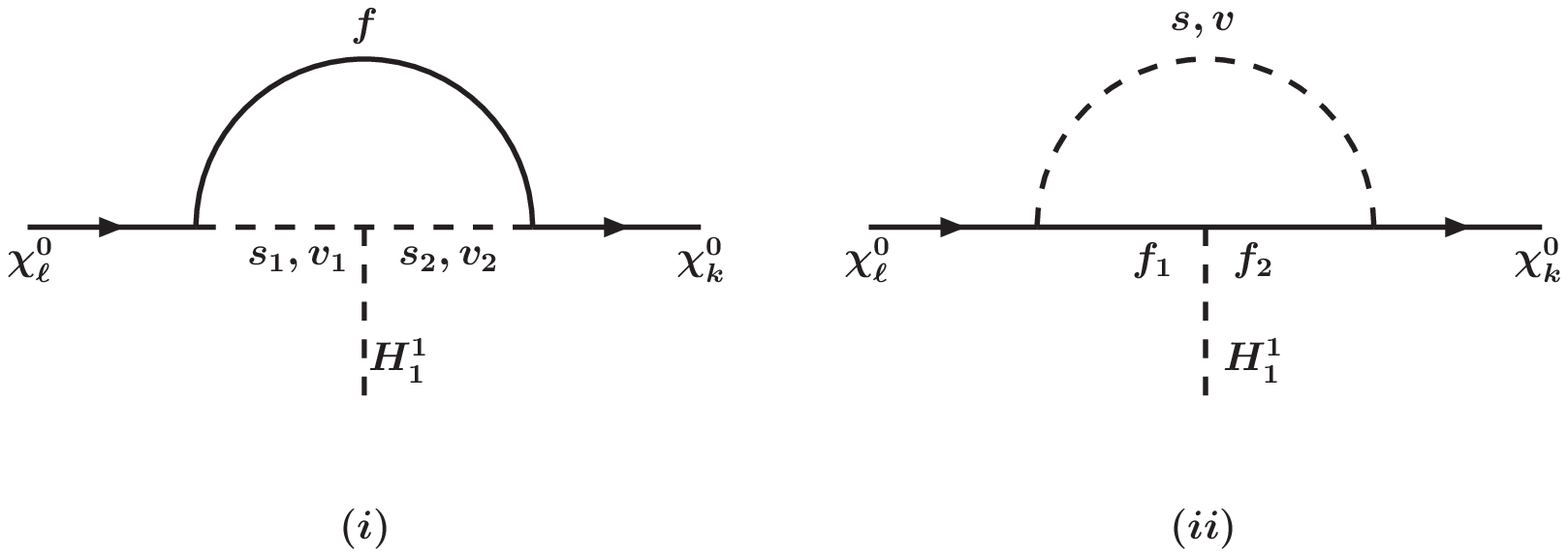}
\caption{Set of diagrams contributing to 
radiative corrections 
$\delta \theta_{k{\ell}}$ and $\Delta \tau_{k{\ell}}$.
 (i): (a) $s_1=\tilde{t}_{j}^{*}$, $s_2=\tilde{t}_{i}^{*}$, $f=t$; (b) $s_1=\tilde{b}_{j}^{*}$, $s_2=\tilde{b}_{i}^{*}$, $f=b$; (c)
$s_1=H^0_n$, $s_2=H^0_m$, $f=\chi^0_i$; (d) $v_1=Z^0$,
$v_2=Z^0$, $f=\chi^{0}_i$; (e) $s_1=H^-$, $s_2=H^-$,
 $f=\chi^{+}_i$; (f) $v_1=W^-$, $v_2=W^-$, $f=\chi^{+}_i$;
(g)$s_1=\tilde{\tau}_i^{*}$, $s_2=\tilde{\tau}_j^{*}$,
 $f=\tau$.
(ii): (a) $f_1=t$, $f_2=t$, $s=\tilde{t}_{j}^{*}$;
(b) $f_1=b$, $f_2=b$, $s=\tilde{b}_{j}^{*}$;
(c)$f_1=\chi^0_i$, $f_2=\chi^0_j$, $v=Z^0$;
(d) $f_1=\chi^0_i$, $f_2=\chi^0_j$, $s=H^0_n$;
(e) $f_1=\chi^+_i$, $f_2=\chi^+_j$, $s=H^-$;
(f) $f_1=\chi^+_i$, $f_2=\chi^+_j$, $v=W^-$;
(g) $f_1=\tau$, $f_2=\tau$, $s=\tilde{\tau}_{i}^{*}$.}
\label{epsfig1}
\end{figure}

\begin{figure}[t]
\hspace*{-0.8in}
\centering
\includegraphics[width=18cm,height=18cm]{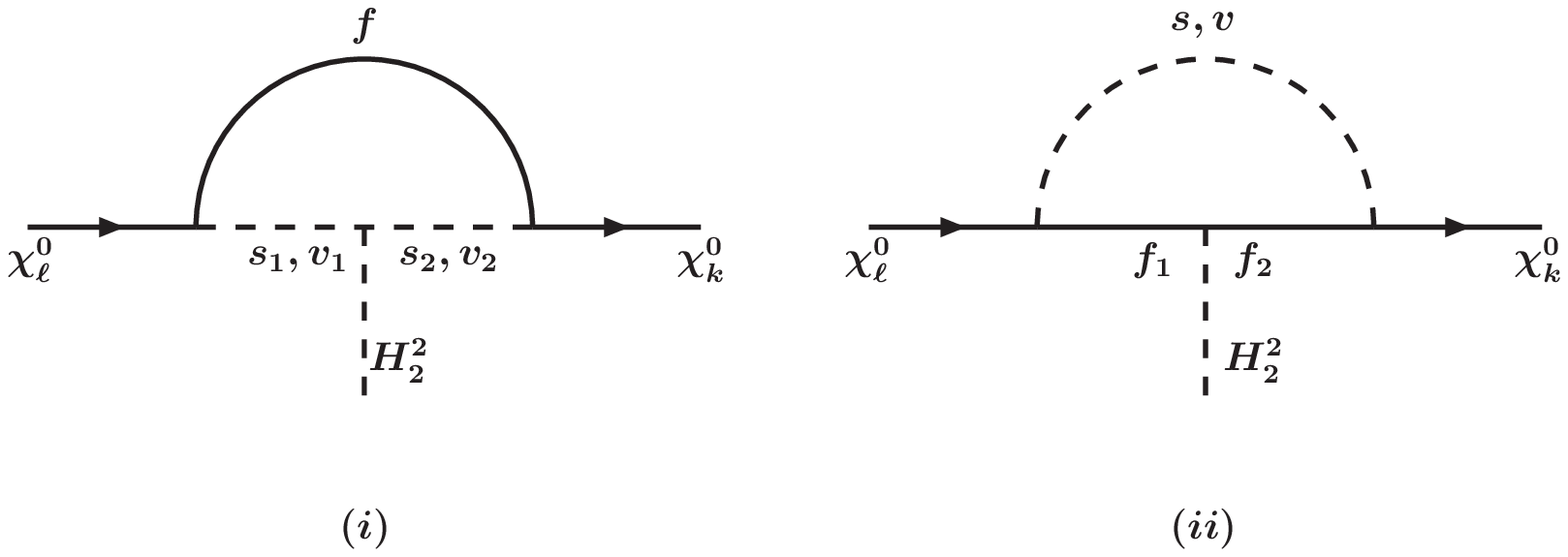}
\caption{Set of diagrams contributing to 
radiative corrections 
$\Delta \theta_{k{\ell}}$ and $\delta \tau_{k{\ell}}$.
(i): (a) $s_1=\tilde{t}_{j}^{*}$, $s_2=\tilde{t}_{i}^{*}$, $f=t$; (b) $s_1=\tilde{b}_{j}^{*}$, $s_2=\tilde{b}_{i}^{*}$, $f=b$; (c)
$s_1=H^0_n$, $s_2=H^0_m$, $f=\chi^0_i$; (d) $v_1=Z^0$,
$v_2=Z^0$, $f=\chi^{0}_i$; (e) $s_1=H^-$, $s_2=H^-$,
 $f=\chi^{+}_i$; (f) $v_1=W^-$, $v_2=W^-$, $f=\chi^{+}_i$;
(g)$s_1=\tilde{\tau}_i^{*}$, $s_2=\tilde{\tau}_j^{*}$,
 $f=\tau$.
(ii): (a) $f_1=t$, $f_2=t$, $s=\tilde{t}_{j}^{*}$;
(b) $f_1=b$, $f_2=b$, $s=\tilde{b}_{j}^{*}$;
(c)$f_1=\chi^0_i$, $f_2=\chi^0_j$, $v=Z^0$;
(d) $f_1=\chi^0_i$, $f_2=\chi^0_j$, $s=H^0_n$;
(e) $f_1=\chi^+_i$, $f_2=\chi^+_j$, $s=H^-$;
(f) $f_1=\chi^+_i$, $f_2=\chi^+_j$, $v=W^-$;
(g) $f_1=\tau$, $f_2=\tau$, $s=\tilde{\tau}_{i}^{*}$.}
\label{epsfig2}
\end{figure}

\begin{figure}[t]
\hspace*{-0.8in}
\centering
\includegraphics[width=18cm,height=18cm]{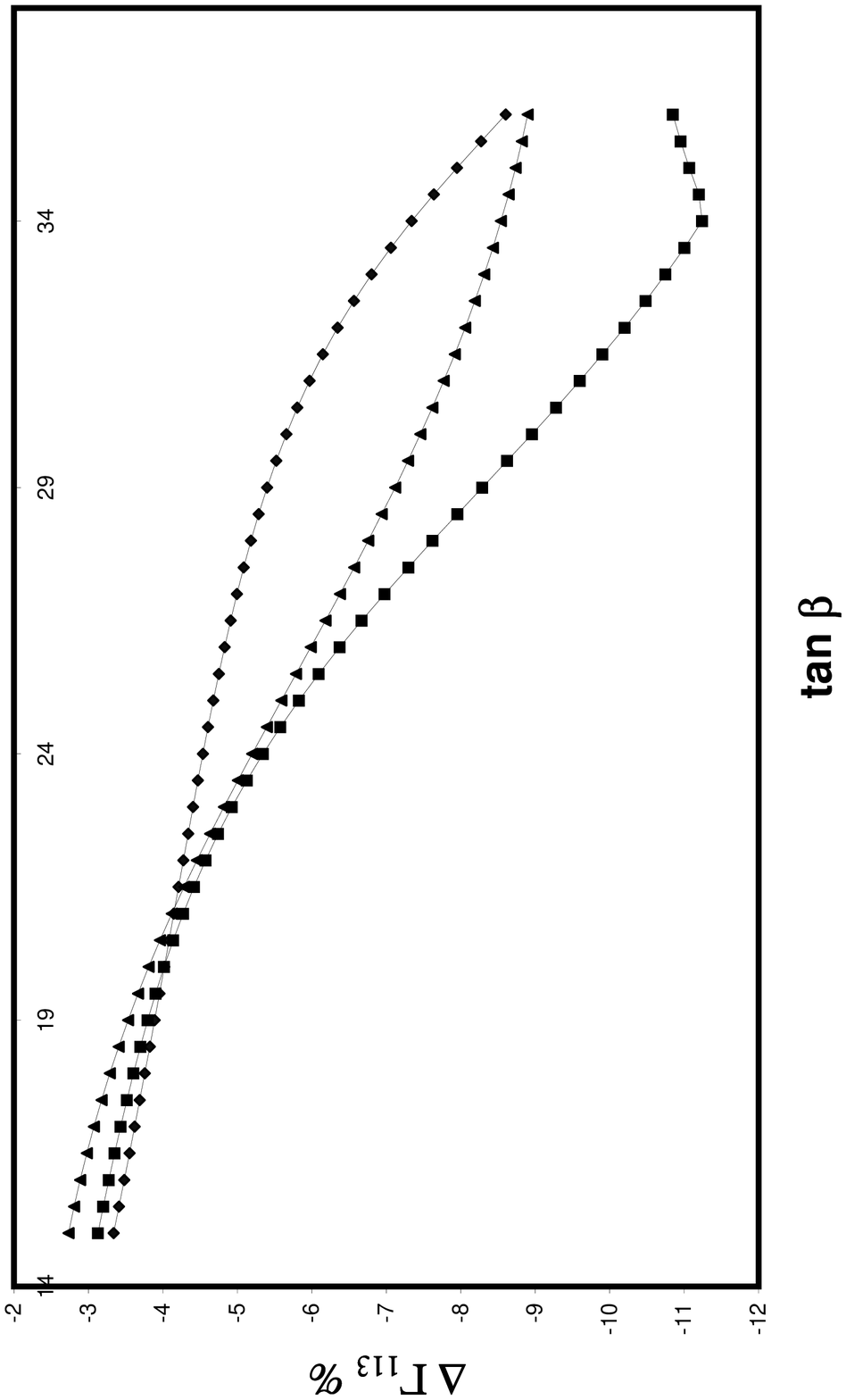}
\caption{$\tan\beta$ dependence of $\Delta \Gamma_1\to \chi_1^0\chi_3^0$.
The curves in ascending order of the absolute value 
at $\tan\beta=15$ 
correspond to $\theta_{\mu}=0.7$, $0.2$ and $0.0$ rad. The input is
 $m_0=300$ GeV,
$m_{1/2}=100$ GeV,
 $|A_0|=100$ GeV,
 $\xi_1=0.5$ (rad), $\xi_2=0.6$ (rad), $\xi_3=0.7$ (rad) and $\alpha_{0}=2.0$ (rad).}
\label{epsfig7a}
\end{figure}

\begin{figure}[t]
\hspace*{-0.8in}
\centering
\includegraphics[width=18cm,height=18cm]{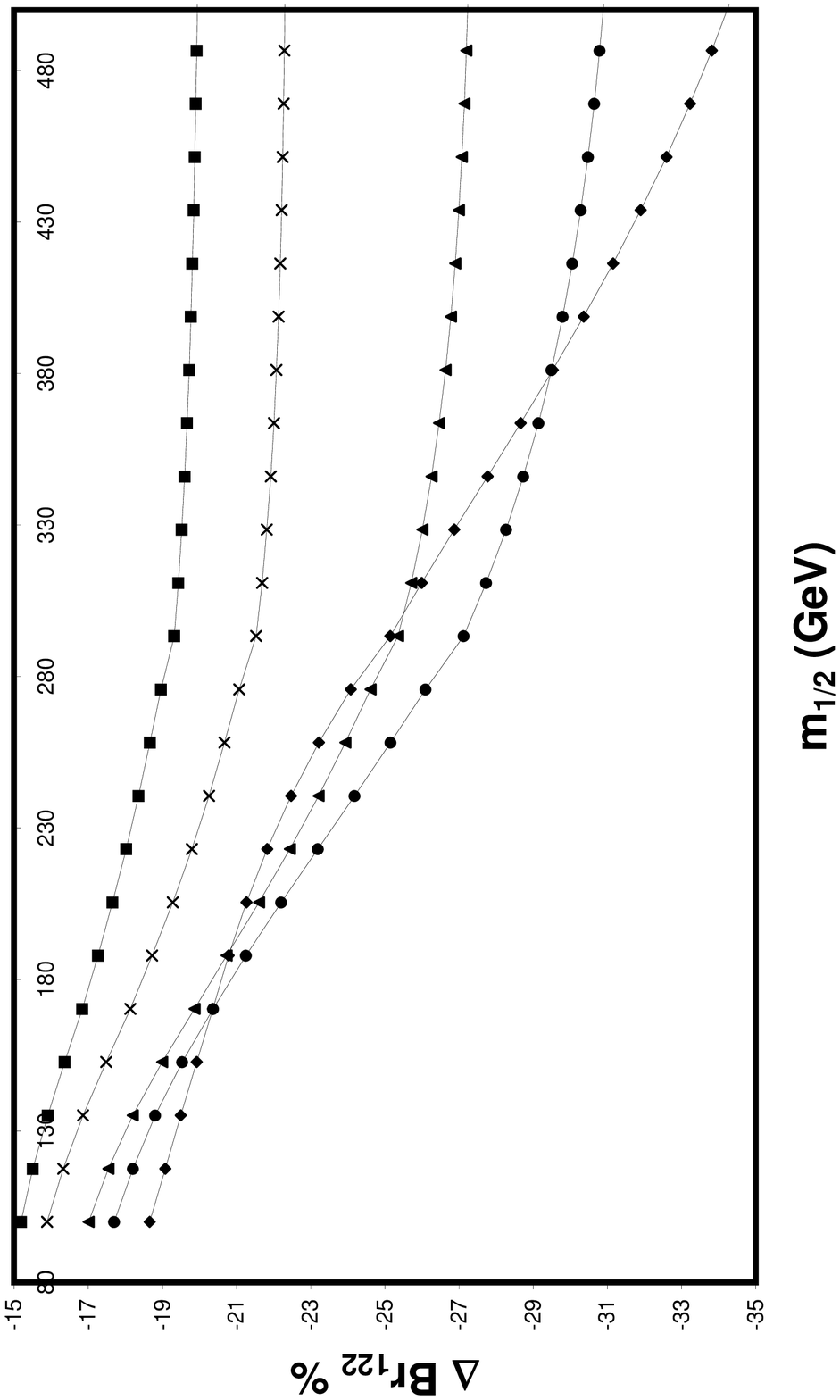}
\caption{$m_{1/2}$ dependence of $\Delta Br_1\to \chi_2^0\chi_2^0$.
The curves in ascending order of the absolute value 
at $m_{1/2}=100$ (GeV)
correspond to $\theta_{\mu}=1.2$, $1.0$, $0.7$, $0.5$ and $0.2$ rad. The input is
$\tan\beta=20$,
 $m_0=500$ GeV,
 $|A_0|=250$ GeV,
 $\xi_1=0.4$ (rad), $\xi_2=0.5$ (rad), $\xi_3=0.6$ (rad) and $\alpha_{0}=0.8$ (rad).}
\label{epsfig7b}
\end{figure}

\begin{figure}[t]
\hspace*{-0.8in}
\centering
\includegraphics[width=18cm,height=18cm]{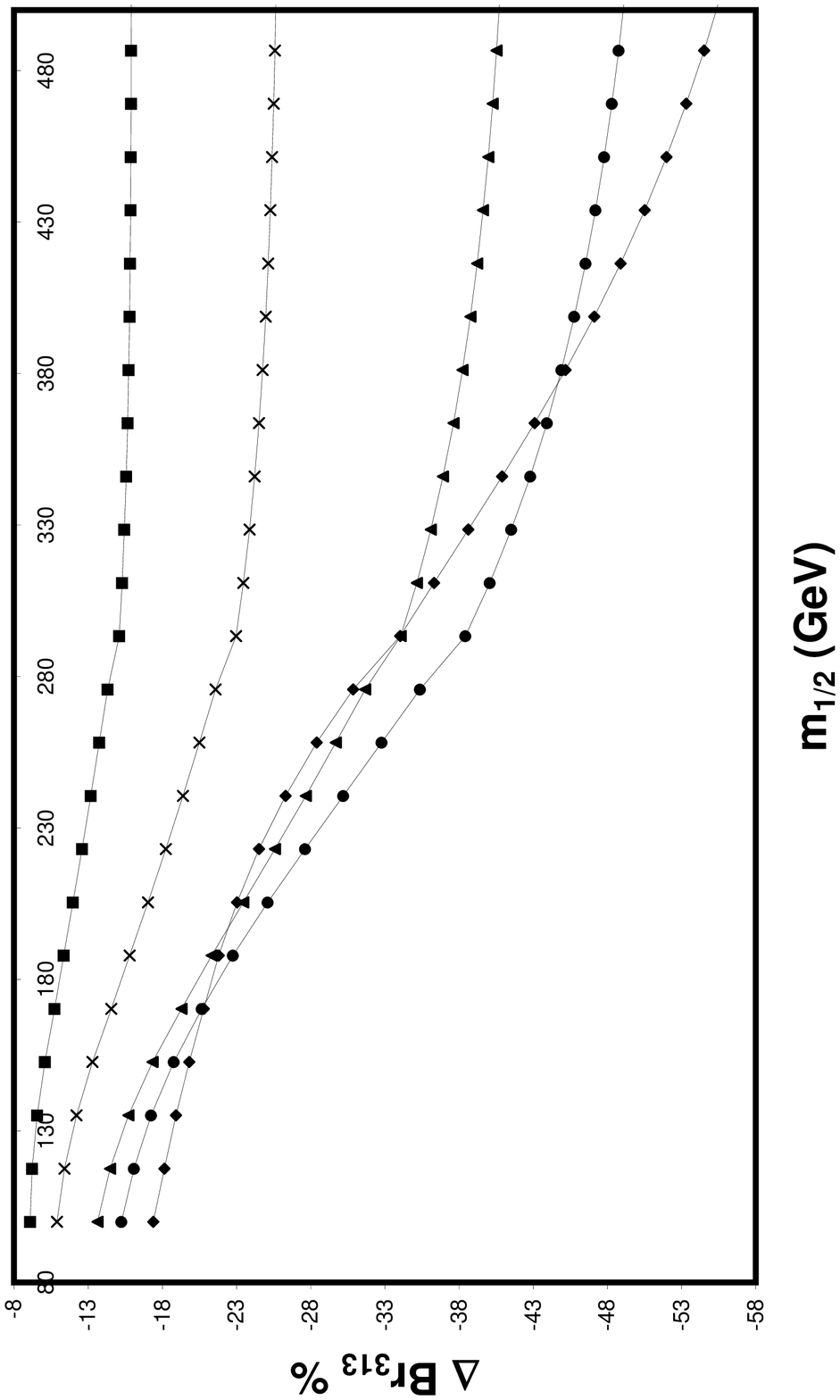}
\caption{$m_{1/2}$ dependence of $\Delta Br_3\to \chi_1^0\chi_3^0$.
The curves in ascending order of the absolute value
at $m_{1/2}=100$ (GeV)
correspond to $\theta_{\mu}=1.2$, $1.0$, $0.7$, $0.5$ and $0.2$ rad. The input is
$\tan\beta=20$,
 $m_0=500$ GeV,
 $|A_0|=250$ GeV,
 $\xi_1=0.4$ (rad), $\xi_2=0.5$ (rad), $\xi_3=0.6$ (rad) and $\alpha_{0}=0.8$ (rad).}
\label{epsfig7c}
\end{figure}

\begin{figure}[t]
\hspace*{-0.8in}
\centering
\includegraphics[width=18cm,height=18cm]{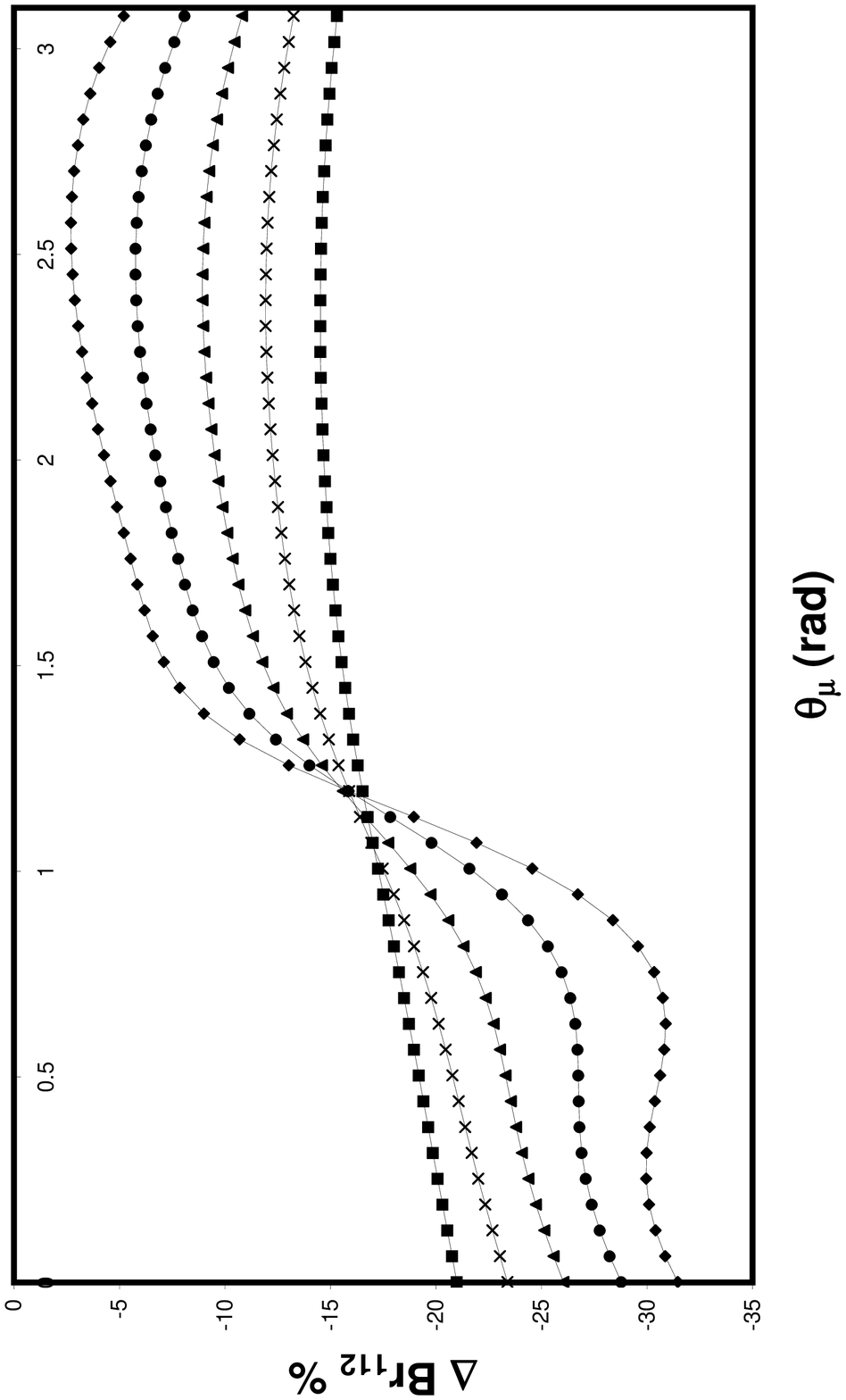}
\caption{$\theta_{\mu}$ dependence of $\Delta Br_1\to \chi_1^0\chi_2^0$.
The curves in descending order of the absolute value
at $\theta_{\mu}=0.0$ (rad)
correspond to $\tan\beta=40$, $35$, $30$, $25$ and $20$. The input is
$m_{1/2}=150$ GeV,
 $m_0=600$ GeV,
 $|A_0|=600$ GeV,
 $\xi_1=0.3$ (rad), $\xi_2=0.4$ (rad), $\xi_3=0.5$ (rad) and $\alpha_{0}=0.5$ (rad).}
\label{epsfig7d}
\end{figure}

\begin{figure}[t]
\hspace*{-0.8in}
\centering
\includegraphics[width=18cm,height=18cm]{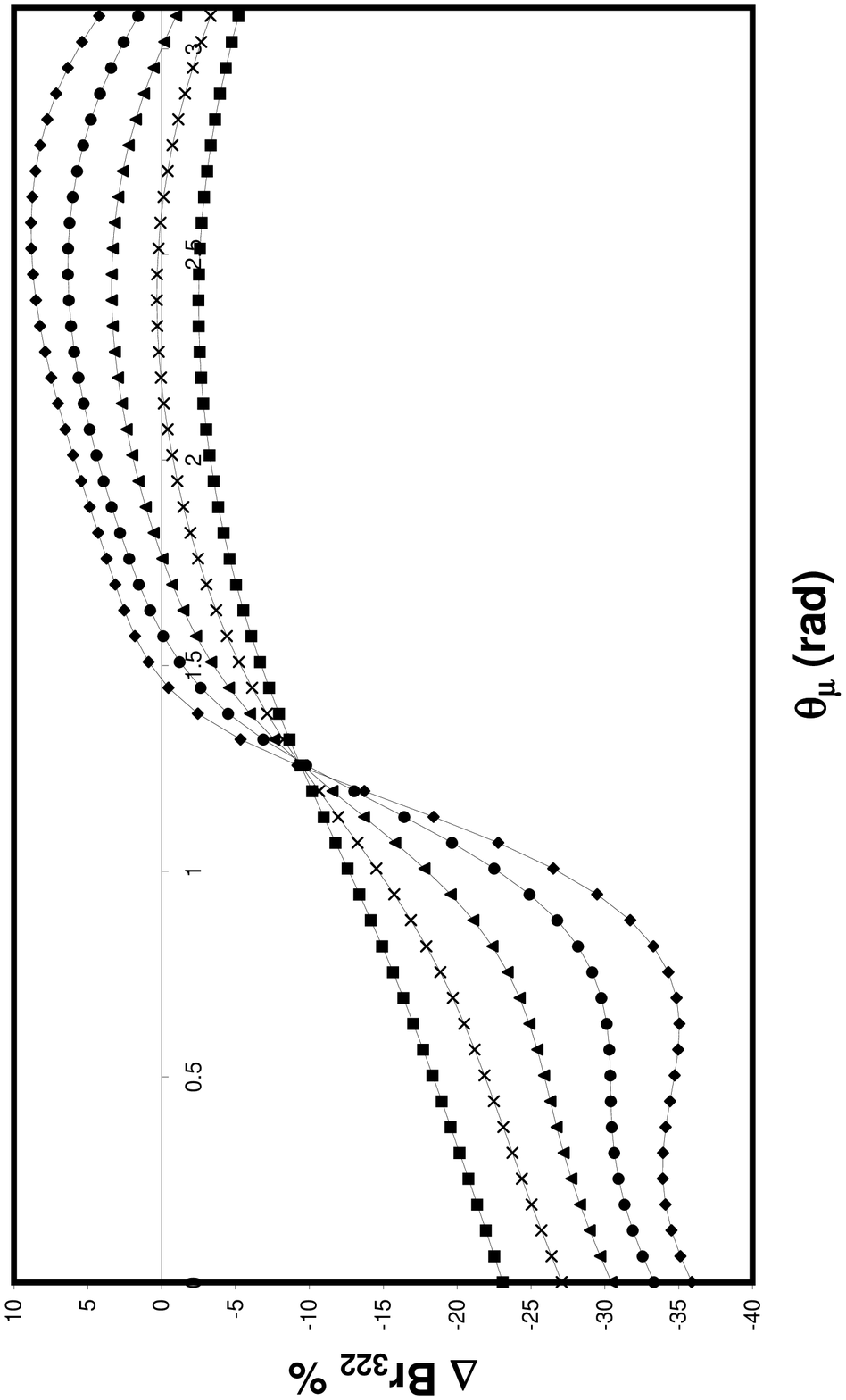}
\caption{$\theta_{\mu}$ dependence of $\Delta Br_3\to \chi_2^0\chi_2^0$.
The curves in descending order of the absolute value
at $\theta_{\mu}=0.0$ (rad)
correspond to $\tan\beta=40$, $35$, $30$, $25$ and $20$. The input is
$m_{1/2}=150$ GeV,
 $m_0=600$ GeV,
 $|A_0|=600$ GeV,
 $\xi_1=0.3$ (rad), $\xi_2=0.4$ (rad), $\xi_3=0.5$ (rad) and $\alpha_{0}=0.5$ (rad).}
\label{epsfig7e}
\end{figure}

\begin{figure}[t]
\hspace*{-0.8in}
\centering
\includegraphics[width=18cm,height=18cm]{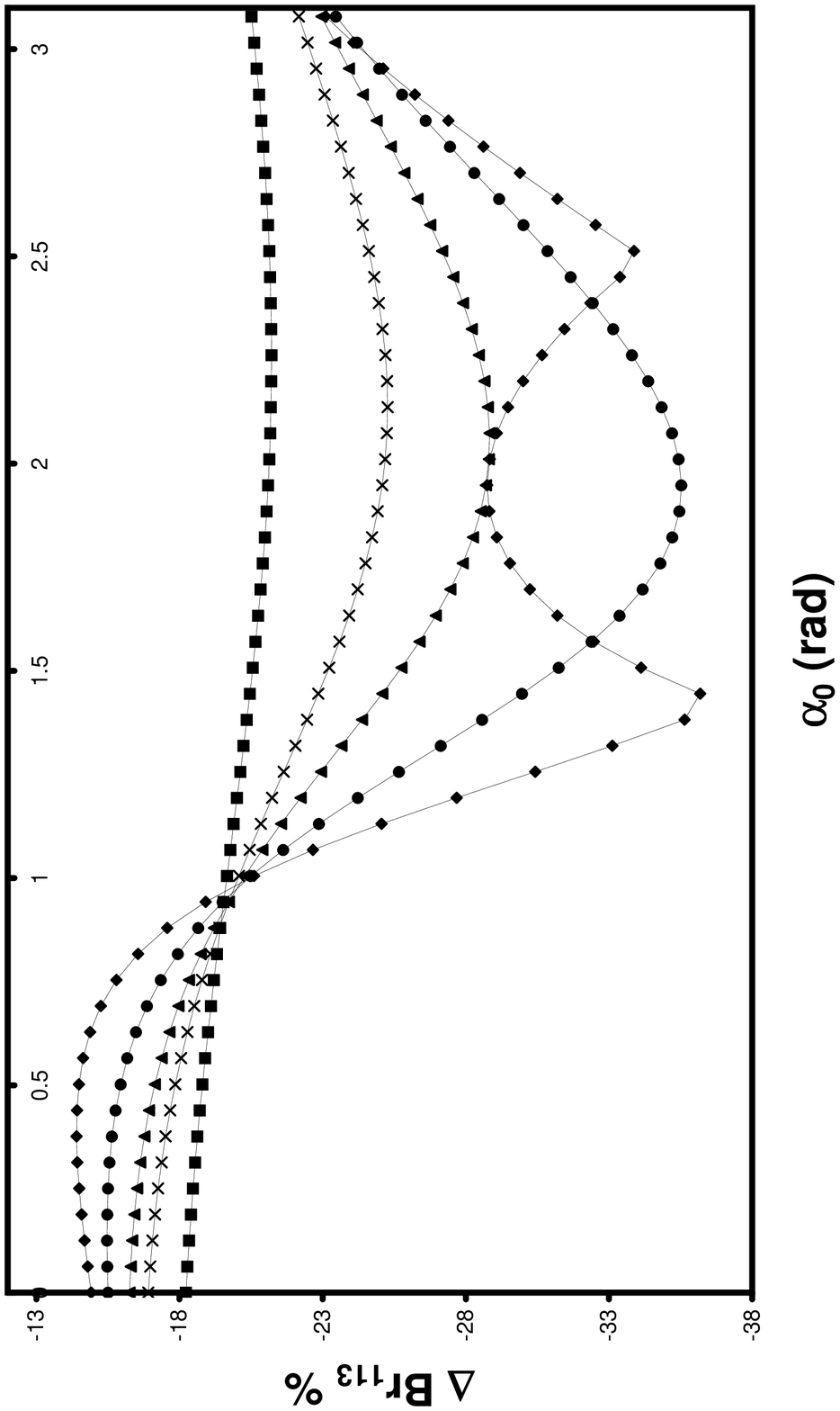}
\caption{$\alpha_{0}$ dependence of $\Delta Br_1\to \chi_1^0\chi_3^0$.
The curves in descending order of the absolute value
at $\alpha_{0}=0.0$ (rad)
correspond to $|A_0|=100$, $250$, $350$, $500$ and $650$ GeV. The input is
$\tan\beta=30$,
$m_{1/2}=150$ GeV,
 $m_0=500$ GeV,
 $\xi_1=0.5$ (rad), $\xi_2=0.6$ (rad), $\xi_3=0.7$ (rad) and $\theta_{\mu}=1.0$ (rad).}
\label{epsfig7f}
\end{figure}

\begin{figure}[t]
\hspace*{-0.8in}
\centering
\includegraphics[width=18cm,height=18cm]{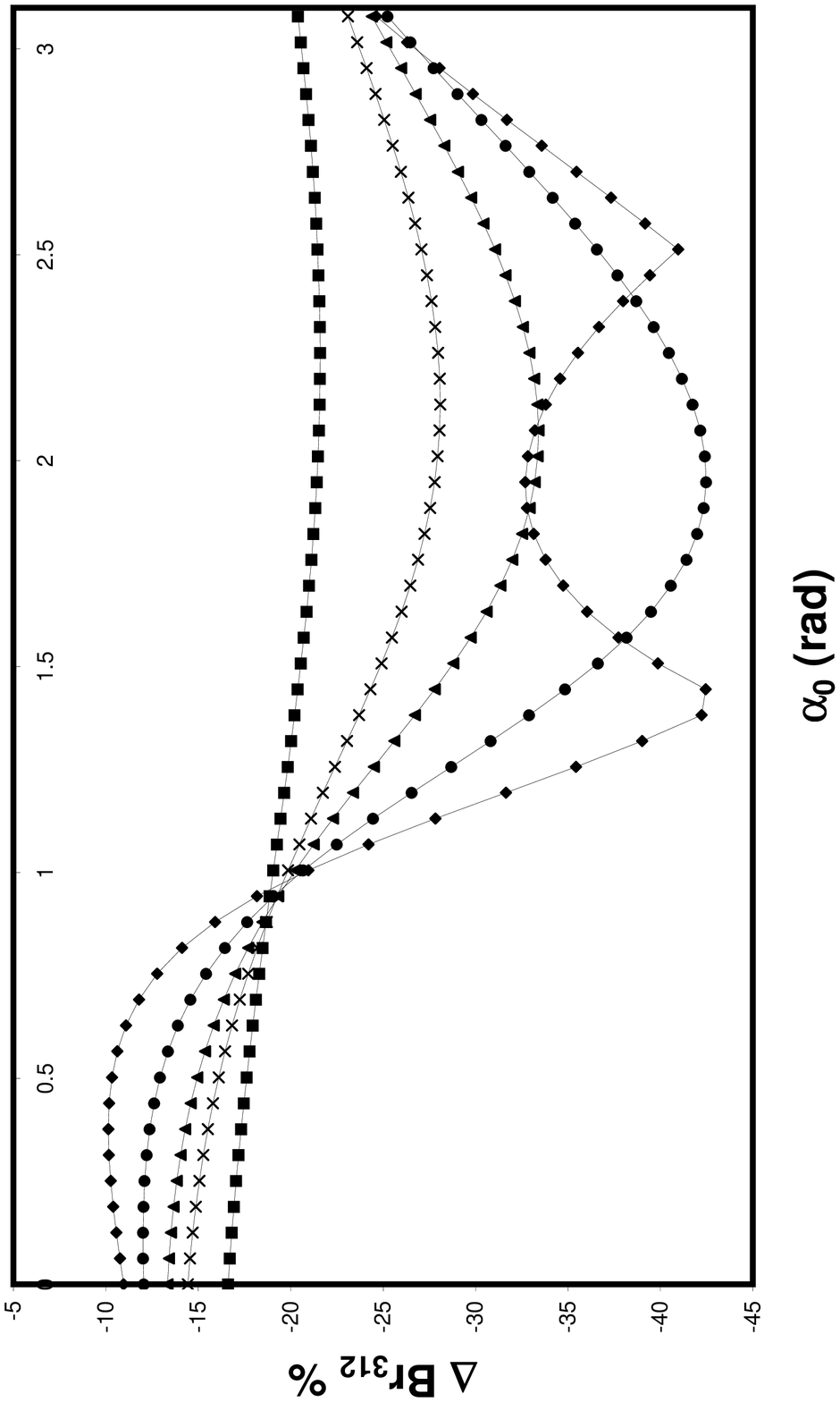}
\caption{$\alpha_{0}$ dependence of $\Delta Br_3\to \chi_1^0\chi_2^0$.
The curves in descending order of the absolute value
at $\theta_{\mu}=0.0$ (rad)
correspond to $|A_0|=100$, $250$, $350$, $500$ and $650$ GeV. The input is
$\tan\beta=30$,
$m_{1/2}=150$ GeV,
 $m_0=500$ GeV,
 $\xi_1=0.5$ (rad), $\xi_2=0.6$ (rad), $\xi_3=0.7$ (rad) and $\theta_{\mu}=1.0$ (rad).}
\label{epsfig7g}
\end{figure}

\begin{figure}[t]
\hspace*{-0.8in}
\centering
\includegraphics[width=18cm,height=18cm]{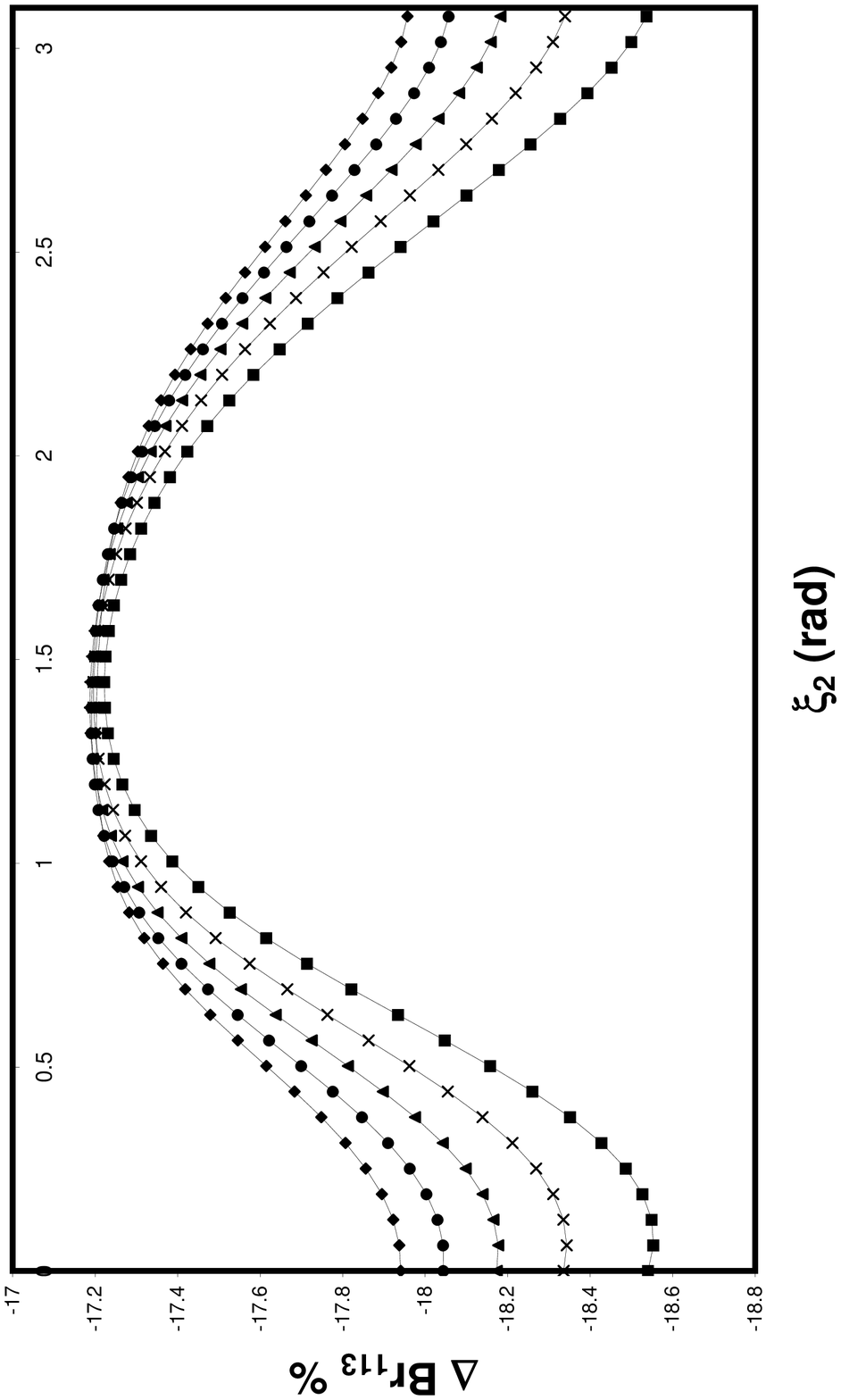}
\caption{$\xi_2$ dependence of $\Delta Br_1\to \chi_1^0\chi_3^0$.
The curves in descending order of the absolute value
at $\xi_2=0.0$ (rad)
correspond to $m_0=650$, $700$, $750$, $800$ and $850$ GeV. The input is
$\tan\beta=20$,
$m_{1/2}=200$ GeV,
 $|A_0|=350$ GeV,
 $\xi_1=0.4$ (rad), $\xi_3=0.6$ (rad), $\alpha_{0}=0.8$ (rad) and $\theta_{\mu}=1.0$ (rad).}
\label{epsfig7h}
\end{figure}

\begin{figure}[t]
\hspace*{-0.8in}
\centering
\includegraphics[width=18cm,height=18cm]{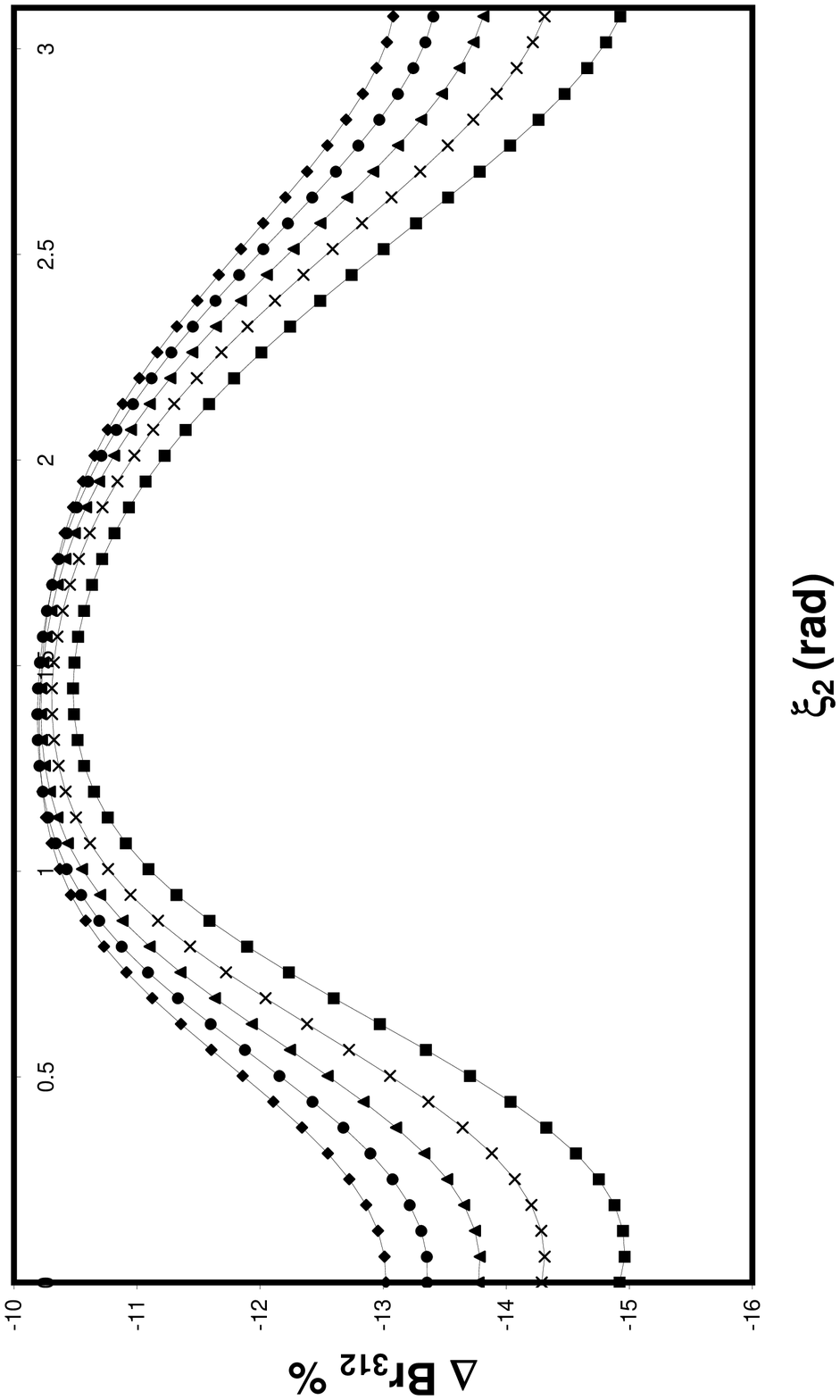}
\caption{$\xi_2$ dependence of $\Delta Br_3\to \chi_1^0\chi_2^0$.
The curves in descending order of the absolute value 
at $\xi_2=0.0$ (rad)
correspond to $m_0=650$, $700$, $750$, $800$ and $850$ GeV. The input is
$\tan\beta=20$,
$m_{1/2}=200$ GeV,
 $|A_0=350$ GeV,
 $\xi_1=0.4$ (rad), $\xi_3=0.6$ (rad), $\alpha_{0}=0.8$ (rad) and $\theta_{\mu}=1.0$ (rad).}
\label{epsfig7i}
\end{figure}

\end{document}